\documentclass[aps, prl,twocolumn,longbibliography,floatfix,aps,nobibnotes]{revtex4-2}

\usepackage{color}
\usepackage{dcolumn}
\usepackage{graphicx}

\usepackage{amsmath,amsfonts,amssymb,amsthm}
\usepackage{mathtools}

\usepackage{bbold}
\usepackage{dsfont}

\usepackage{soul}

\usepackage{fancyhdr}
\usepackage{multirow}
\usepackage{bm}
\usepackage{dsfont}
\usepackage{cancel}

\usepackage{xifthen}

\usepackage{float}
\graphicspath{{"./graphs/"}{"./scheme/"}}

\usepackage{graphicx}
\usepackage[usenames,dvipsnames,dvipsnames,svgnames]{xcolor}
\usepackage[colorlinks,bookmarks=false,
linkcolor=blue,
urlcolor=blue,
citecolor=blue
]{hyperref}

\usepackage{xspace}

\def\dissip#1{\mathcal{D}\big[#1\big]}

\def\commut#1#2{\lbrack #1,#2 \rbrack}

\newcommand{\q}[1]{``#1"}
\usepackage{physics}

\usepackage{xifthen}

\def\op#1{\hat{#1}}
\renewcommand{\ao}[1][]{%
	\ifthenelse{\equal{#1}{}}{\ensuremath{\op{a}}}{\ensuremath{\op{#1}}}%
}

\newcommand{\co}[1][]{%
	\ifthenelse{\equal{#1}{}}{\ensuremath{{{}\op{a}^{\dagger}}}}{\ensuremath{{{}\op{#1}^{\dagger}}}}%
}

\makeatletter
\newcommand*{\transpose}{\bgroup\@transpose}
\newcommand*{\@transpose}[1][0]{\mathpalette\@@transpose{#1}\egroup}
\newcommand*{\@@transpose}[2]{\setbox0=\hbox{\m@th$#1\mkern-#2mu\intercal$}\raise\dp0\box0}
\makeatother

\usepackage{array,calc}

\usepackage[english]{babel}
\usepackage{pgfplots} 
\pgfplotsset{compat=1.18}
\usetikzlibrary{plotmarks}

\usepackage{scalerel}

\usepackage[normalem]{ulem}

\def\ext{0}
\usepackage{ifthen}
\usepackage{xifthen}
\newcommand{\comments}[1]{
	\ifthenelse{\equal{\ext}{1}}{\textcolor{blue}{#1}}{}
}
\def\extFuture{1}
\definecolor{darkgreen}{RGB}{0,140,0}

\graphicspath{{"../graphs/"}{"../images/"}{"../scheme/"}{"figures/"}
}
\begin{document}

\title{Floquet expansion by counting pump photons}
\author{Kilian Seibold}
\thanks{These authors contributed equally to this work.}
\affiliation{Department of Physics, University of Konstanz, 78464 Konstanz, Germany}
\author{Orjan Ameye}
\thanks{These authors contributed equally to this work.}
\affiliation{Department of Physics, University of Konstanz, 78464 Konstanz, Germany}
\author{Oded Zilberberg}
\affiliation{Department of Physics, University of Konstanz, 78464 Konstanz, Germany}
\date{\today}

\begin{abstract}
Periodically-driven systems engender a rich competition between the time scales of the drives and those of the system, leading to a limited ability to describe the system in full.
We present a framework for the description of interacting bosonic driven systems via a Floquet expansion on top of a quantization that ``counts'' the drive photons, and provide compelling arguments for the superior performance of our method relative to standard Floquet approaches. Crucially, our approach extends beyond the rotating wave approximation and addresses the long-standing issue of mismatch between the quantum Floquet formalism and its classical counterpart. We, furthermore, pinpoint key corrections to the positions of multiphoton resonances, which are commonly used in the calibration and operation of qubit architectures. 
\end{abstract}

\maketitle


\section{Introduction}

Floquet engineering stands at the forefront of contemporary research both in experiment and theory. It has the goal of generating, characterizing and controlling quantum states of matter in few- and many-body systems~\cite{Eckardt2017, Oka2019, Rudner2020, Holthaus2015, Jotzu2014, Aidelsburger2013, Lohse2018, Lohse2015}.
It relies on using time-periodic external fields, and has seen a wide array of applications; including the generation, manipulation and control of quantum states~\cite{Sato2020, Pieplow2019, Bai2021, Qiu2022, Chen2015, Restrepo2016, Viola1999, Liu2023}, 
the engineering of tailored material properties~\cite{Wang2020, Oka2019, Weitenberg2021, Wang2022a}, enhancing quantum sensing and metrology~\cite{eichler2018,heugel2019quantum,Bai2023}, the creation and characterization of topological states of matter~\cite{Rudner2020a, Harper2020, Kitagawa2010, Lindner2011}, and the creation of synthetic gauge fields~\cite{Wang2020,del2023dynamical}. Floquet engineering similarly stands at the heart of cat qubit architectures~\cite{Grimm2020} and coherent control of ultrafast dynamics~\cite{Mattes2024, Budden2021}.
%

Analyzing periodically-driven quantum systems is technically challenging because the time-dependence breaks energy conservation, leads to out-of-equilibrium conditions, and renormalizes the system's parameters (AC-Stark shifts)~\cite{BlochSiegert1940}. 
To address this, Floquet's theorem is commonly used, where fast oscillations are averaged out and the system is effectively described by a time-independent model with slow variables. 
High-frequency perturbative techniques such as van Vleck~\cite{Eckardt2015}, Floquet-Magnus~\cite{Casas2001, Blanes2009}, or Brillouin-Wigner~\cite{Mikami2016} expansions are then employed to approximate the resulting time-independent effective model.
At first order, these expansions boil down to the widely-employed rotating wave approximation (RWA)~\cite{Eckardt2017, 
Eckardt2015, Bukov2015, Goldman2014},
%
which provides a good approximate description when the driving field is weak and closely resonant with the system's energy scales~\cite{Ann2021}. 
%
In the classical realm, a similar method called the Krylov–Bogolyubov (KB) or averaging method~\cite{Zilberberg2023, Krylov1949, Bogoliubov1961, Sanders2007, Guckenheimer2013, Oliveira2017, Neu1980, Holmes1981} is often used to deal with time-dependent differential equations.

The compatibility between quantum and classical limits, as $\hbar \rightarrow 0$, is crucial for an unified theory spanning macroscopic and microscopic scales. However, quantum Floquet-based methods exhibit discrepancies with the Ehrenfest's theorem~\cite{Ford1996}, and the RWA often fails in strongly-driven or strongly-detuned systems~\cite{Joergensen2022,Laucht2016,Ann2021}.
While numerous works tackled the breakdown of the RWA in spin systems~\cite{Sharaby2010,Zheng2008,Zhang2015}, or spins coupled to harmonic oscillators~\cite{Zueco2009,Zeuch2020,Wang2021,Gan2010,He2012,Hausinger2008}, recent focus extends to bosonic systems with applications in quantum information processing~\cite{Nourmandipour2021, Cortinas2024}, quantum sensing~\cite{Zhang2021} and simulation~\cite{Schindler2013,Sandholzer2019},
optomechanical cooling~\cite{Liu2013,Marquardt2008}, frequency combs~\cite{Weng2022,Herr2012,Chembo2016,Lugiato1987,del2023limit}, and realization of single-photon sources~\cite{Flayac2017,Ling2023,Chen2022}.
At the center of this broad range of activities is the driven anharmonic oscillator, a.k.a. the Duffing (Kerr) oscillator or single-site driven Bose-Hubbard model, where RWA fails to accurately capture key effects, such as multi-photon resonances (MPRs) and phase transitions in the system~\cite{Xiao2023, Biondi2017, Casteels2017}.

In this work, we present a formalism for high-frequency expansions in bosonic systems that resolves the longstanding challenge of reconciling quantum and classical formalisms. We leverage a recently proposed concept emphasizing the role of the operator basis in bosonic mode quantization~\cite{Kosata2022}. Thus, we redefine the quantum operators within a basis tailored to anticipate the system's response at the driving frequency. Crucially, our systematic expansion reconciles order-by-order the quantum-to-classical treatments, and quickly converges towards the exact stationary response. Furthermore, our method outperforms existing approaches in describing resonant quantum effects. We provide experimental protocols for validating our findings. Our general formalism can lead to a precise depiction of a plethora of driven open quantum many-body systems. 

\begin{figure}[t!]
	\centering
	\includegraphics[width=1.0\linewidth]{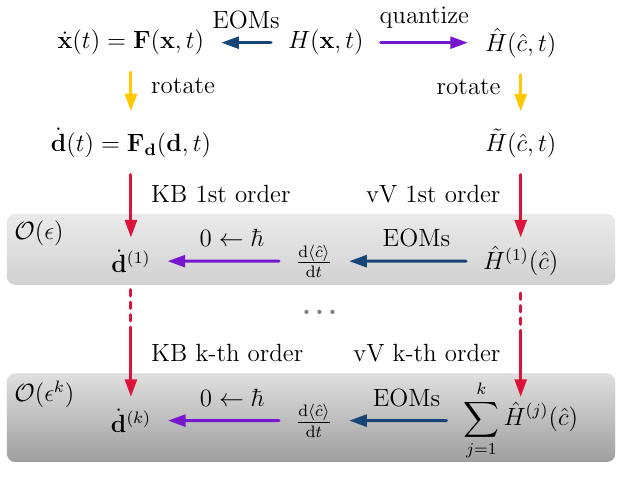}
	\caption{\textit{Flowchart on how to analyze time-dependent Hamiltonians $H(x,t)$} [cf. example in Eq. \eqref{eq: Classical Hamiltonian KO}]. In the classical limit (left side), one derives Hamilton's equations of motion (EOMs, left black arrow) and employs the Krylov-Bogoliubov
averaging method [cf. Eqs. \eqref{eq: slow-flow KB2}], which engenders a systematic perturbation theory with increasing orders (down red arrows). In the quantum limit (right side), the coordinates are quantized [cf. Eq. \eqref{eq: ladder operators a}, right violet arrow], and e.g., a quantum van Vleck perturbation expansion is employed [cf. Eqs. \eqref{eq: H effective generic}-\eqref{eq: F hamiltonian c ops}, down red arrows]. Both treatments rely on a rotation relative to the drive frequency. Using second-quantized operators that count the photons of the drive, our methods obtains a matching quantum-to-classical limit, when employing mean-field Heisenberg's equation of motion and taking the semiclassical limit ($\hbar\rightarrow 0$, left violet arrow).}
	\label{fig: diagram}
\end{figure}

We develop a framework in both the classical and quantum formalisms [see Fig.~\ref{fig: diagram}] for the analysis of periodic time-dependent Hamiltonians $H(\vb{x},t)$, where $\vb{x}\coloneqq(x,p)$ are phase space coordinates.
As an example, we consider a single driven Duffing oscillator~\cite{Lifshitz2008}
\begin{equation}\label{eq: Classical Hamiltonian KO}
H(\vb{x}, t)=\frac{{p}^2}{2m}+\frac{m\omega_0^2}{2}x^2+\frac{\alpha}{4} {x}^4-F\cos{(\omega t)}{x}\:,
\end{equation}
where $m$ is the mass of the oscillator mode, $\omega_0$ its natural/bare frequency, $\alpha$ the Duffing nonlinearity, 
and $F$ and $\omega$ are the strength and frequency of the external drive, respectively. Without loss of generality, we take $\alpha>0$~\cite{Zilberberg2023}.

\emph{The classical formalism.---}%
To treat driven nonlinear classical systems, we employ the KB averaging method, cf.~left branch of Fig.~\ref{fig: diagram}.
Starting from Hamilton's equations of motion (EOMs) $\dv{\vb{x}}{t}=\dot{\vb{x}} = \vb{F}(\vb{x}, t)$, we first move to a frame rotating at the frequency of the drive $\omega$ by assuming $x=U(t) \cos(\omega t) + V(t) \sin(\omega t)$ with slowly time-evolving $U(t)$ and $V(t)$. This rotating ansatz is catered towards treating nearly-resonant driving of the system, and encapsulates the lore that driven linear systems tend to respond at the drive frequency.
In this rotating frame, the equations become $\dot{\vb{d}}=\vb{F}_{\vb{d}}(\vb{d},t)$ for the coordinates $\mathbf{d}(t) \coloneqq (U(t),V(t))$.
These EOMs remain explicitly time-dependent with a period $T\coloneqq\frac{\pi}{\omega}$, but can be 
approximated by time-independent EOMs via a perturbative expansion
$\vb{d}(t)=\sum_{i=0}\epsilon^i \vb{d}_i(t)$ with $\epsilon\ll 1$ and by solving for the (stroboscopic) slow-flow quadratures $\vb{d}_0\coloneqq (u,v)$. The latter contains the remaining time-dependence on timescales much longer than $T$.
To lowest order in the expansion, we have $\vb{d}_0 \coloneqq[\,\vb{F}_{\vb{d}}\,]_\mathrm{av}$, where the brackets represent time averaging $[\,\cdot\,]_\mathrm{av}\equiv\frac{1}{T}\int_0^{T} \cdot \, \dd{t}$. Higher-order corrections can be systematically generated~\cite{Bogoliubov1961,Zilberberg2023,supmat}.
Thus, the lowest-order KB description of Eq.~\eqref{eq: Classical Hamiltonian KO} reads
\begin{align}\label{eq: slow-flow KB2}
\displaystyle\renewcommand\arraystretch{1.5}
   \dot{\vb{d}}_0=
   \mqty(\dot{u}\\\dot{v})
    =\mqty(
    \frac{\omega^2-\omega_0^2}{2\omega}v-\frac{3\alpha X^2}{8\omega}v
    \\
    -\frac{\omega^2-\omega_0^2}{2\omega}u+\frac{3\alpha X^2}{8\omega}u+\frac{F}{2\omega}
    ),
\end{align}
where $X^2\coloneqq u^2+v^2$ is the amplitude of the stroboscopic motion~\cite{supmat}. For the expansion to be valid, the terms 
    $\alpha X^2/\left(m\omega^2\right),
    \ \abs{(\omega^2-\omega_0^2)}/\omega^2,
    \ \sqrt{\left(\alpha F^2\right)/\left(m^3\omega^6\right)} \sim \epsilon$ must be small~\cite{supmat}.

%
\emph{The quantum formalism.---}%
The classical time-dependent Hamiltonian $H(\vb{x}, t)$ can be second-quantized using bosonic ladder operators, $\hat{c}$,  that describe the annihilation (lowering) of a bosonic particle (photon number) with frequency $\omega_c$
\begin{equation}\label{eq: ladder operators a}
	\hat{x}=\sqrt{\dfrac{\hbar}{2 m \omega_c}}
     \,(\hat{c}^\dagger+\hat{c}) \qq{and} \hat{p}=i \sqrt{\dfrac{m \hbar \omega_c}{2}}\,(\hat{c}^\dagger-\hat{c})\;.
\end{equation}
Thus, we obtain a quantum Hamiltonian $\hat{H}(\hat{c}, t)$, cf.~right branch of Fig.~\ref{fig: diagram}. Commonly, we use (canonical) ladder operators $\hat{c}\equiv\hat{a}$, where $\hat{a}$ is the lowering operator relative to the bare frequency $\omega_c\equiv \omega_0$. The nondriven harmonic oscillator is diagonal in this operator basis with energies that depend on the photon number $\hbar \omega_0 \co\ao$. In other words, by counting the system's photons, we diagonalize the nondriven harmonic part of the system.
The standard procedure to treat time-periodic quantum systems is analoguous to the classical one [right branch of Fig.~\ref{fig: diagram}].
Micromotion is separated from the stroboscopic time evolution by moving to a rotating frame using the unitary operator $\mathcal{U}_{c}(t)=\exp(-i \omega t \hat{c}^{\dagger} \hat{c})$ and obtain the rotating frame Hamiltonian:
$\tilde{H}(\hat{c},t) := \mathcal{U}_{c}^{\dagger} \, \hat{H}(t) \, \mathcal{U}_{c}-i \hbar \mathcal{U}_{c}^{\dagger} \, \partial_t \,\mathcal{U}_{c}$.
In this frame, the time-dependent (counter-rotating) terms in $\tilde{H}(\hat{c},t)$ exhibit rapid oscillations with frequencies higher than the driving frequency $\omega$.

\footnotetext[5]{Here, we follow naming convention similar the one use in Ref.~\cite{Eckardt2015}, i.e., the effective Hamiltonian is the time-independent Hamiltonian which represent the stroboscopic dynamics of the system independent of the Floquet gauge --- in contrast to the Floquet Hamiltonian.}

To obtain an effective time-independent description, we employ a Floquet expansion in orders of the fast counter-rotating frequency $1/2\omega$. Specifically, we use van Vleck (vV) degenerate perturbation theory, yielding an effective Hamiltonian $\hat{H}_\mathrm{eff}$~\cite{Note5}. The rotated Hamiltonian $\tilde{H}(\hat{c},t)$ is decomposed into Fourier components $\tilde{H}_l = \left[\tilde{H} \exp(-i l \omega t)\right]_\mathrm{av}$ such that the effective Hamiltonian can be written as~\cite{Eckardt2015}
\begin{equation}
    \label{eq: H effective generic}
    \tilde{H}_\mathrm{eff} = \sum_{\nu=1} \tilde{H}^{(\nu)}.
\end{equation}
The first-order term, $\tilde{H}^{(1)} = \tilde{H}_0  $, is also known as the RWA, whereas the next term is computed by $\tilde{H}^{(2)} = \sum_{l\neq0} \tilde{H}_l \tilde{H}_{-l}/(l \hbar \omega)$. 
Applying this procedure to our example~\eqref{eq: Classical Hamiltonian KO}, we find the RWA effective Hamiltonian
\begin{equation}\label{eq: F hamiltonian c ops}
\tilde{H}_{\mathrm{eff},c} = \hbar(-\Delta_c\ + U_c)\, \co[c]\ao[c] +  \dfrac{\hbar U_c}{2}\, \co[c]\co[c] \ao[c]\ao[c] -  \hbar F_c \,(\ao[c] + \co[c])\;,
\end{equation}
%
with detuning $\Delta_c$, nonlinearity $U_c$, and pump strength $F_c$, which are assumed to be small compared with $\hbar\omega$; these are the same validity conditions as in the KB method \cite{supmat}. Using the canonical $\hat{a}$-operators, we have $\Delta_a = \omega-\omega_0$, $U_a = (3\alpha \hbar)/(8 m^2 \omega_0 ^2)$, and $F_a=F/(2\sqrt{2m\omega_0\hbar})$.

\emph{The quantum-to-classical limit.---}%
Both the quantum and classical approaches outlined above provide an effective stroboscopic description of the system.
In both methods, we rotate the system at the driving frequency $\omega$ and average-out the fast dynamics in a perturbative manner. 
Therefore, we expect that both yield the same result when taking the quantum-to-classical limit. 
This limit is obtained analytically by (i) deriving the averaged Heisenberg equations of motion
$
\frac{d}{dt}\langle\hat{c}\rangle 
= \frac{1}{i\hbar}\langle[\hat{c},\hat{H}]\rangle
+ \langle\frac{\partial \hat{c}}{\partial t}\rangle
$,
for the observable $\hat{c}$, (ii) applying a mean-field approximation such that $\hat{c}\to\langle\hat{c}\rangle$, and (iii) taking the limit $\hbar \to 0$. Carrying out this procedure to the example~\eqref{eq: F hamiltonian c ops}, quantized using the standard ladder operator $\hat{a}$, we do not retrieve the same equations as the classical slow-flow EOMs~\eqref{eq: slow-flow KB2}. This discrepancy manifests at any order, which is the first result of this work~\cite{supmat}.

\begin{figure}[t!]
	\centering
	\includegraphics[width=\linewidth]{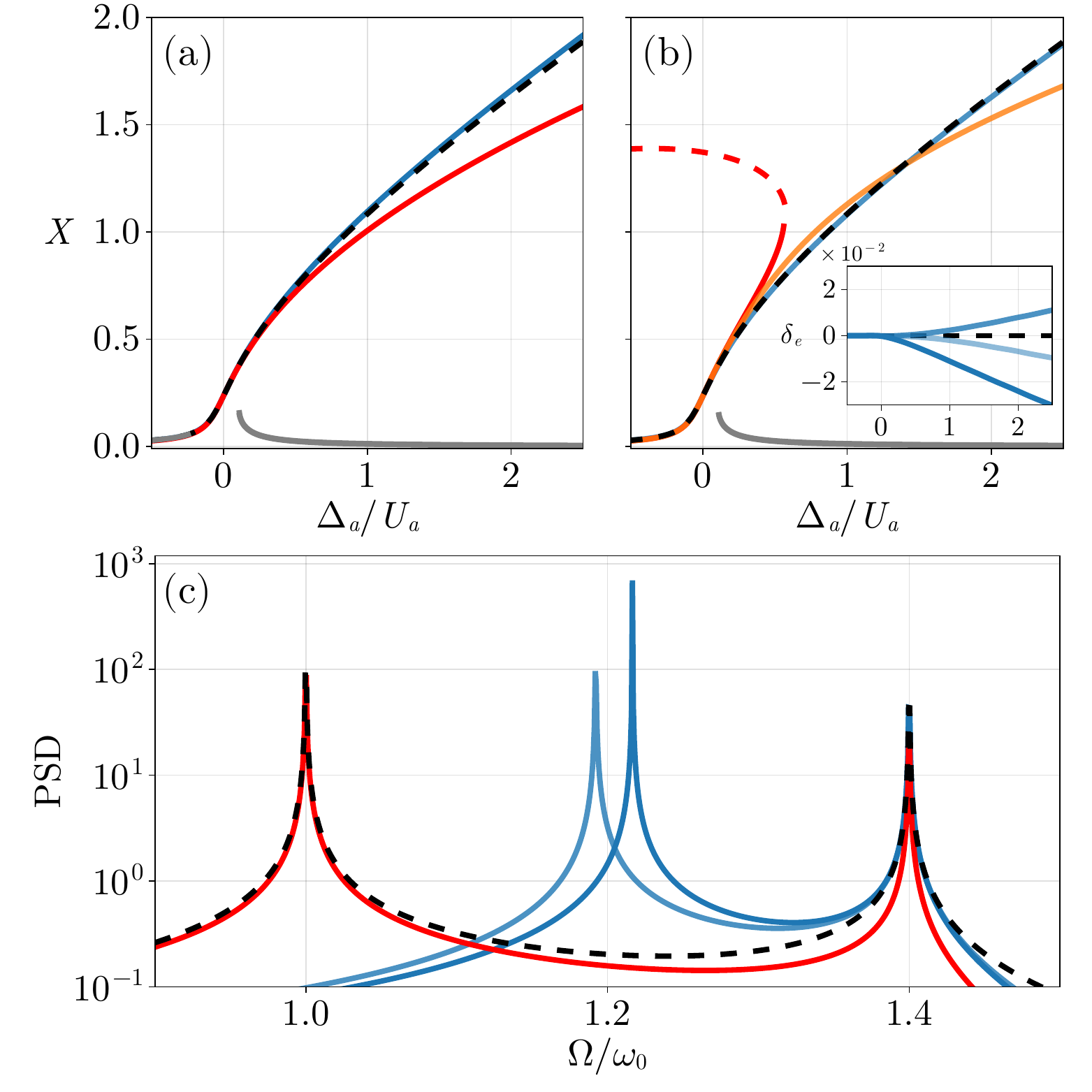}
	\caption{\textit{Comparison between the expansion methods relative to the classical numerical solution}.
        (a) and (b) Classical stationary amplitude, $X = \sqrt{u^2+v^2}$, of the driven Duffing~\eqref{eq: Classical Hamiltonian KO} at the drive frequency $\omega$ as a function of detuning $\Delta_a/U_a$, with $m=\omega_0=1$ and $F_a/U_a= 10^{-2}$. We compare the exact numerical solution (dashed, black line) with (a) the first-order van Vleck expansion [cf.~Eq.~\eqref{eq: F hamiltonian c ops}] in the $\hat{a}$- (solid red) and $\hat{b}$- (blue solid) operator basis [cf.~Eq.~\eqref{eq: ladder operators a}]; the latter is equivalent to the KB result~\eqref{eq: slow-flow KB2}. (b) Comparison with higher-order corrections [cf.~Eq.~\eqref{eq: H effective generic}]: second and third order in $\ao$-basis (red solid) and second order in $\ao[b]$-basis (orange solid). Inset: Absolute error $\delta_e$ compared to the exact of the first, second and third order (progressively lighter blue) in the $\hat{b}$-basis. A small dissipation term $\gamma \dot{x}$ is added to enforce numerical convergence with $\gamma/\omega_0=2.5\times10^{-3}$. 
        (c) Power Spectral Density (PSD) versus response frequency $\Omega$ of the driven harmonic oscillator, cf.~Eq.~\eqref{eq: Classical Hamiltonian KO} with $m=\omega_0=1$, $\Delta_a/\omega_0=0.4$, $\alpha=0$, and $F_a/\omega_0=3.5\times10^{-3}$. The exact analytical solution (with marginally-broadened Dirac deltas)~\cite{supmat} (dashed black) is compared with the first-order van Vleck in the $\hat{a}$ (solid blue) and $\hat{b}$ (solid blue) basis. 
  }
    \label{fig: fig2}
\end{figure}

To find out which of the EOMs perform better in the classical limit, we compare them to a numerical \q{experiment}, see Figs.~\ref{fig: fig2}(a) and (b). We time-evolve the classical time-dependent Hamilton's EOMs, obtained from Eq.~\eqref{eq: Classical Hamiltonian KO}, until a stationary oscillation is reached, and plot the amplitude $\abs{X}$ of the stationary oscillation harmonic at $\omega$. As a function of detuning, the system exhibits two possible stationary phases (low- and high-amplitude response) that have a coexistence region. 
We focus here on the high-amplitude branch.
Whereas already to lowest-order the stationary solution ($\dot{u}=\dot{v}=0$) of the classical KB equations~\eqref{eq: slow-flow KB2} coincide with the numerical result, we observe that the stationary ($d\langle\ao[c]\rangle/dt = 0$) vV method fails to converge up to third order [cf. red and orange lines Fig.~\ref{fig: fig2}(a-b)], despite the fact that we are within the validity bounds of the perturbative approximation. Furthermore, the vV second order correction breaks down, such that, at large detuning, we even obtain an spurious instability of the solution. Higher-order corrections reconcile the instability, but the stationary amplitude deviates significantly from the exact numerical solution. Note that this discrepancy is related to the fact that the standard RWA relative to $\mathcal{U}_{a}(t)$ breaks Ehrenfest's theorem~\cite{Ford1996, Kosata2022}. The KB method, instead, shows consistent convergence towards the numerical exact solution [see inset].

\emph{Counting the photons of the drive.---}%
To resolve this longstanding problem in the quantum Floquet expansion, we are inspired by the classical KB approach, and postulate that the system responds mostly at the driving frequency~\cite{Kosata2022}. To this end, we write the quantum Hamiltonian using the ladder operators $\hat{c}\equiv\hat{b}$, where we ``count'' the photons of the drive, i.e., $\omega_c=\omega$. Furthermore, we move to a rotating frame described with $\mathcal{U}_{b}(t)$, and perform the vV expansion in this frame.
In our case study~\eqref{eq: Classical Hamiltonian KO}-\eqref{eq: F hamiltonian c ops}, this change of basis manifests in the effective Hamiltonian~\eqref{eq: F hamiltonian c ops} as a rescaling of the coefficients, namely, we have $\Delta_b = (\omega^2 - \omega_0^2)/(2  \omega)$, nonlinearity $U_b = U_a  (\omega_0 / \omega)^2$, and pump strength $F_b = F_a \sqrt{\omega_0 / \omega}$. Note that these operators act on a different Fock basis than the $\op{a}$-operators and a Bogoliubov transformation is required when moving between counting of the system to counting of the drive photons~\cite{supmat,Note10}.
%
%
Crucially, performing the expansion within the rotating frame determined with the $\hat{b}$-operator restores the quantum-to-classical limit~\cite{supmat}.
In other words, we obtain exactly the same EOMs by taking the semiclassical limit of vV expansion, as those obtained by the classical KB approach. Moreover, the correspondence is maintained up to the third order, with higher orders pending further investigation.
This is the main result of this work.
\footnotetext[10]{Applying the quantum-to-classical limit to the unitary rotating frame transformation $\mathcal{U}_b(t)$ yields the same rotation transformation used in the classical framework. This is not the case when working in the $a$-basis and the rotating frame $\mathcal{U}_a(t)$~\cite{supmat}. 
Specifically, we can write $\hat{a} = \hat{S}^\dagger\hat{b}\hat{S}$, with $\hat{S}=\exp(z/2(\co^2+\ao^2))$, where the parameter $z$ controls the symplectic transformation and is a function of $\omega_0$ and $\omega$. Here, $\ao = \mu\ao[b]-\nu\co[b]$, with $\mu=1/2(\sqrt{\omega/\omega_0}+\sqrt{\omega_0/\omega})$, and $\nu=1/2(\sqrt{\omega/\omega_0}-\sqrt{\omega_0/\omega})$. The parameter $z$ is defined by $\mu=\cosh(|z|)$ and $\nu=z/|z|\sinh(|z|)$.}




%

\emph{Closed system considerations.---}%
We showed that the stationary response of the system is better described based on counting the pump photons. To this end, we assumed the existence of a time-independent solution in the appropriately chosen rotating frame, aided by the inclusion of a minuscule damping term $\gamma \dot{x}$ with $\gamma\ll \omega_0$ in our numerical simulations to facilitate convergence, i.e., we ensure \q{causality}. Convergence to a stationary solution occurs over a relatively extended period $t\gg 1/\gamma$, during which initial boundary conditions become negligible. The main response is at the drive frequency $\omega$.

In a fully closed system ($\gamma\equiv 0$), however, even the response of the driven harmonic oscillator ($\alpha=0$) exhibits two distinct peaks: one at the bare frequency $\omega_0$ and the other at the driving frequency $\omega$, see Fig.~\ref{fig: fig2}(c). The former relies on initial conditions, and the latter characterizes the response to the drive~\cite{Zilberberg2023,supmat}. Comparing this case to the standard RWA in the $\hat{a}$-operators basis yields approximate descriptions of both peaks in terms of frequency, yet their amplitudes are inaccurate. In contrast, using the $\hat{b}$-basis provides a precise amplitude description at $\omega$ already at first order, albeit with inaccuracies in both frequency and amplitude regarding the peak at $\omega_0$. The latter approximation is corrected via higher-order squeezing terms in the expansion~\cite{supmat}. Consequently, the effectiveness of our $\hat{b}$-basis approach varies with system details and initial conditions at short times, but we anticipate improved performance over long (stationary) periods, assuming  causality.

\emph{Quantum observables.---}%
Moving beyond the semiclassical limit, we turn to explore the performance of our theory in the quantum realm.
A notable quantum effect manifesting in our example~\eqref{eq: F hamiltonian c ops} are multiphoton resonances (MPRs)~\cite{Anikin2021, Ruiz2023}. This feature is characterized by a resonant increase in the photon number inside the cavity, and has no classical counterpart. It holds utmost importance since the spacing (in detuning) between neighboring MPRs is used in the calibration of the photon-photon interaction strength in superconducting qubit experiments~\cite{Winkel2020}.
Based on the RWA model in the $\hat{a}$-basis, the MPRs can be solved analytically~\cite{Drummond1980, Bartolo2016, Roberts2020}. Here, we generalize this expression to any arbitrary choice of the operator basis using Eq.~\eqref{eq: F hamiltonian c ops}. 
For example the position of where the MPRs arises is given by
\begin{equation}
\Delta_c/U_c = (n-1)/2\, , \qquad n=1,2, \dots \,.
    \label{eq: formula multiphoton resonances closed system}
\end{equation}
%
%
Crucially, whereas in the standard $\ao$-basis, equidistant peaks are observed in $\Delta_a$, we predict that in the $\ao[b]$-basis, the peaks become denser with increasing detuning.

\begin{figure}[t!]
	\centering
 \includegraphics[width=\linewidth]{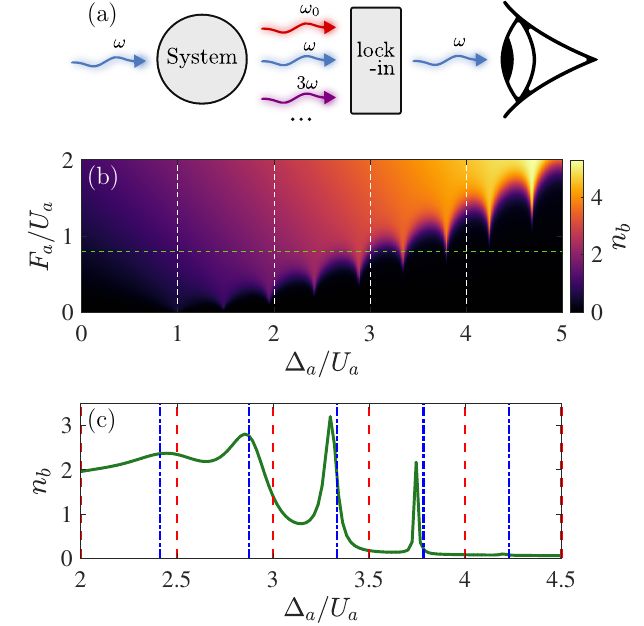}
	\caption{\textit{Comparison in the quantum realm}.
(a) Sketch of the system [cf.~Eqs.~\eqref{eq: Classical Hamiltonian KO}] driven by photons of frequency $\omega$, and emitting photons at $\omega_0$, $\omega$, and wave-mixed harmonics, e.g., at $3\omega$. A heterodyne detector (lock-in) filters out only photons at $\omega$. (b) The numerically calculated [cf.~Eq.~\eqref{eq: quantum master equation}] stationary photon number $n_b$ as a function of detuning $\Delta_a$ and pump power $F_a$ for $\tilde{H}_{\mathrm{eff},b}$. The numerical evolution is performed in properly truncated Hilbert space, ensuring convergence by varying the cutoff. Vertical dashed line marks $U_a/\omega_0 = 10^{-2}$ and $F_a/U_a = 0.8$, corresponding to (c). (c) Comparison between the numerical exact solution (green line) and the predictions from Eq.~\eqref{eq: formula multiphoton resonances closed system} in $\op{a}$ (red, dashed lines) and $\op{b}$ basis (blue, dot-dashed lines).
}
	\label{fig: Wigner fct open F=3.5}
\end{figure}

\emph{MPR prediction vs. Lindblad exact time evolution.---} To test which approach fares better in the quantum realm, we perform a numerical experiment akin to the classical case in Fig.~\ref{fig: fig2}. We evolve the system's density matrix, $\op{\rho}$, using the time-dependent exact Hamiltonian~\eqref{eq: Classical Hamiltonian KO}. Motivated by heterodyne (Lock-in) measurements~\cite{Meade1982}, 
where the system is driven with photons at frequency $\omega$, and the lock-in detects the response with emitted photons at $\omega$ [Fig.~\ref{fig: Wigner fct open F=3.5}(a)], we (i) work in the $\op{b}$ basis representation,
and (ii) evolve using the rotated Hamiltonian $\tilde{H}(\hat{b},t)$. 
Furthermore, we are interested in the MPRs in the long-time limit. 
To reach a stationary state, we add a small dissipation term
%
in the form of a photon-loss Lindblad superoperator $\mathcal{D}\big[\ao[b]\big]\op{\rho} = \ao[b]\op{\rho}\ao[b]^{\dagger}-1/2(\ao[b]^{\dagger}\ao[b]\op{\rho}+ \op{\rho}\ao[b]^{\dagger}\ao[b])$, and evolve the system using the Lindblad quantum master equation~\cite{Breuer2007,Gardiner2000}:
\begin{equation}
\dfrac{d\op{\rho}}{dt}= -\frac{i}{\hbar}\commut{\tilde{H}}{\op{\rho}}+ \kappa \dissip{\ao[b]}{\op{\rho}}\,,
\label{eq: quantum master equation}
\end{equation}
where 
$\kappa$ is the dissipation rate. 
%
%
In the long time limit, the system evolves towards a nonequilibrium stationary state, where we calculate the average photon number from the density matrix $\left[\Tr{\op{\rho}(t_{\rm long}) \co[b] \ao[b]}\right]_{\mathrm{av}}$; and present its time-average over several periods. Performing this for $\tilde{H}_{\mathrm{eff},b}$ as a function of detuning $\Delta_a$
and pump power $F_a$ clearly show the MPRs, see Fig.~\ref{fig: Wigner fct open F=3.5}(b).
%
%
Taking a line cut, we can quantitatively compare which expansion, in which operator basis, approximates the exact solution better, see Fig.~\ref{fig: Wigner fct open F=3.5}(c). Crucially, the MPRs obtained from the exact time evolution in the rotating frame do not appear with a constant detuning spacing, and better coincide with the $\op{b}$ and not the $\op{a}$ basis.

%




We introduced a tailored operator basis designed for analyzing periodically-driven systems, offering an enhanced starting point for perturbative approaches. Our alternative second-quantization of the Hamiltonian anticipates the system's response at the driving frequency, significantly enhancing the predictive accuracy of stationary-state outcomes derived from high-frequency expansions, while maintaining simplicity. Moving beyond previous results, we establish an order-by-order reconciliation between the quantum and classical limits of the perturbation theory, along with refined and unified validity bounds for perturbations in both regimes. We furthermore demonstrate that by counting drive photons, improved agreement with exact models is obtained. In the quantum realm, we predict discrepancies with existing models, which are readily observable in circuit QED experiments. Given the prevalence of periodically driven nonlinear systems across various physics disciplines and the broad applicability of our analysis, we anticipate our findings to impact diverse areas of research. Furthermore, the extension of this formalism to other frameworks such as mean field Gross–Pitaevskii equations, quantum cumulant expansion, or phase space methods will be the topic of future investigations.

\begin{acknowledgments}
	We thank A. Leuch, T. Kästli and J. Ko\v{s}ata for insightful input at the early stage of the project, as well as fruitful discussion with S. Benkert, A. Eichler, A. Eckardt, and J. del Pino. We acknowledge funding from the Deutsche Forschungsgemeinschaft (DFG) via project number 449653034 and through SFB1432, as well as the Swiss National Science Foundation (SNSF) through the Sinergia Grant No.~CRSII5\_206008/1.
\end{acknowledgments}


\bibliographystyle{KilianStyle}
\bibliography{RWA_biblio.bib}

\begin{thebibliography}{90}%
\makeatletter
\providecommand \@ifxundefined [1]{%
 \@ifx{#1\undefined}
}%
\providecommand \@ifnum [1]{%
 \ifnum #1\expandafter \@firstoftwo
 \else \expandafter \@secondoftwo
 \fi
}%
\providecommand \@ifx [1]{%
 \ifx #1\expandafter \@firstoftwo
 \else \expandafter \@secondoftwo
 \fi
}%
\providecommand \natexlab [1]{#1}%
\providecommand \enquote  [1]{``#1''}%
\providecommand \bibnamefont  [1]{#1}%
\providecommand \bibfnamefont [1]{#1}%
\providecommand \citenamefont [1]{#1}%
\providecommand \href@noop [0]{\@secondoftwo}%
\providecommand \href [0]{\begingroup \@sanitize@url \@href}%
\providecommand \@href[1]{\@@startlink{#1}\@@href}%
\providecommand \@@href[1]{\endgroup#1\@@endlink}%
\providecommand \@sanitize@url [0]{\catcode `\\12\catcode `\$12\catcode
  `\&12\catcode `\#12\catcode `\^12\catcode `\_12\catcode `\%12\relax}%
\providecommand \@@startlink[1]{}%
\providecommand \@@endlink[0]{}%
\providecommand \url  [0]{\begingroup\@sanitize@url \@url }%
\providecommand \@url [1]{\endgroup\@href {#1}{\urlprefix }}%
\providecommand \urlprefix  [0]{URL }%
\providecommand \Eprint [0]{\href }%
\providecommand \doibase [0]{http://dx.doi.org/}%
\providecommand \selectlanguage [0]{\@gobble}%
\providecommand \bibinfo  [0]{\@secondoftwo}%
\providecommand \bibfield  [0]{\@secondoftwo}%
\providecommand \translation [1]{[#1]}%
\providecommand \BibitemOpen [0]{}%
\providecommand \bibitemStop [0]{}%
\providecommand \bibitemNoStop [0]{.\EOS\space}%
\providecommand \EOS [0]{\spacefactor3000\relax}%
\providecommand \BibitemShut  [1]{\csname bibitem#1\endcsname}%
\let\auto@bib@innerbib\@empty
\bibitem [{\citenamefont {Eckardt}(2017)}]{Eckardt2017}%
  \BibitemOpen
  \bibinfo {author} {A.~Eckardt},\ \emph {\bibinfo {title} {Colloquium: Atomic
  quantum gases in periodically driven optical lattices}},\ \href
  {\doibase10.1103/revmodphys.89.011004} {\bibfield  {journal} {\bibinfo
  {journal} {Reviews of Modern Physics}\ }\textbf {\bibinfo {volume} {89}},\
  \bibinfo {pages} {011004} (\bibinfo {year} {2017})}\BibitemShut {NoStop}%
\bibitem [{\citenamefont {Oka}\ and\ \citenamefont {Kitamura}(2019)}]{Oka2019}%
  \BibitemOpen
  \bibinfo {author} {T.~Oka}\ and\ \bibinfo {author} {S.~Kitamura},\ \emph
  {\bibinfo {title} {Floquet Engineering of Quantum Materials}},\ \href
  {\doibase10.1146/annurev-conmatphys-031218-013423} {\bibfield  {journal}
  {\bibinfo  {journal} {Annual Review of Condensed Matter Physics}\ }\textbf
  {\bibinfo {volume} {10}},\ \bibinfo {pages} {387} (\bibinfo {year}
  {2019})}\BibitemShut {NoStop}%
\bibitem [{\citenamefont {Rudner}\ and\ \citenamefont
  {Lindner}(2020{\natexlab{a}})}]{Rudner2020}%
  \BibitemOpen
  \bibinfo {author} {M.~Rudner}\ and\ \bibinfo {author} {N.~Lindner},\ \emph
  {\bibinfo {title} {The {Floquet} {Engineer}'s {Handbook}}},\ \href
  {https://www.semanticscholar.org/paper/The-Floquet-Engineer%27s-Handbook-Rudner-Lindner/d077a2c15a633a825077d924d5023e1ec14fb837}
  {\bibfield  {journal} {\bibinfo  {journal} {arXiv: Mesoscale and Nanoscale
  Physics}\ } (\bibinfo {year} {2020}{\natexlab{a}})}\BibitemShut {NoStop}%
\bibitem [{\citenamefont {Holthaus}(2015)}]{Holthaus2015}%
  \BibitemOpen
  \bibinfo {author} {M.~Holthaus},\ \emph {\bibinfo {title} {Floquet
  engineering with quasienergy bands of periodically driven optical
  lattices}},\ \href {\doibase10.1088/0953-4075/49/1/013001} {\bibfield
  {journal} {\bibinfo  {journal} {Journal of Physics B: Atomic, Molecular and
  Optical Physics}\ }\textbf {\bibinfo {volume} {49}},\ \bibinfo {pages}
  {013001} (\bibinfo {year} {2015})}\BibitemShut {NoStop}%
\bibitem [{\citenamefont {Jotzu}\ \emph {et~al.}(2014)\citenamefont {Jotzu},
  \citenamefont {Messer}, \citenamefont {Desbuquois}, \citenamefont {Lebrat},
  \citenamefont {Uehlinger}, \citenamefont {Greif},\ and\ \citenamefont
  {Esslinger}}]{Jotzu2014}%
  \BibitemOpen
  \bibinfo {author} {G.~Jotzu}, \bibinfo {author} {M.~Messer}, \bibinfo
  {author} {R.~Desbuquois}, \bibinfo {author} {M.~Lebrat}, \bibinfo {author}
  {T.~Uehlinger}, \bibinfo {author} {D.~Greif}\ and\ \bibinfo {author}
  {T.~Esslinger},\ \emph {\bibinfo {title} {Experimental realization of the
  topological Haldane model with ultracold fermions}},\ \href
  {\doibase10.1038/nature13915} {\bibfield  {journal} {\bibinfo  {journal}
  {Nature}\ }\textbf {\bibinfo {volume} {515}},\ \bibinfo {pages} {237}
  (\bibinfo {year} {2014})}\BibitemShut {NoStop}%
\bibitem [{\citenamefont {Aidelsburger}\ \emph {et~al.}(2013)\citenamefont
  {Aidelsburger}, \citenamefont {Atala}, \citenamefont {Lohse}, \citenamefont
  {Barreiro}, \citenamefont {Paredes},\ and\ \citenamefont
  {Bloch}}]{Aidelsburger2013}%
  \BibitemOpen
  \bibinfo {author} {M.~Aidelsburger}, \bibinfo {author} {M.~Atala}, \bibinfo
  {author} {M.~Lohse}, \bibinfo {author} {J.~T. Barreiro}, \bibinfo {author}
  {B.~Paredes}\ and\ \bibinfo {author} {I.~Bloch},\ \emph {\bibinfo {title}
  {Realization of the Hofstadter Hamiltonian with Ultracold Atoms in Optical
  Lattices}},\ \href {\doibase10.1103/physrevlett.111.185301} {\bibfield
  {journal} {\bibinfo  {journal} {Physical Review Letters}\ }\textbf {\bibinfo
  {volume} {111}},\ \bibinfo {pages} {185301} (\bibinfo {year}
  {2013})}\BibitemShut {NoStop}%
\bibitem [{\citenamefont {Lohse}\ \emph {et~al.}(2018)\citenamefont {Lohse},
  \citenamefont {Schweizer}, \citenamefont {Price}, \citenamefont
  {Zilberberg},\ and\ \citenamefont {Bloch}}]{Lohse2018}%
  \BibitemOpen
  \bibinfo {author} {M.~Lohse}, \bibinfo {author} {C.~Schweizer}, \bibinfo
  {author} {H.~M. Price}, \bibinfo {author} {O.~Zilberberg}\ and\ \bibinfo
  {author} {I.~Bloch},\ \emph {\bibinfo {title} {Exploring 4D quantum Hall
  physics with a 2D topological charge pump}},\ \href
  {\doibase10.1038/nature25000} {\bibfield  {journal} {\bibinfo  {journal}
  {Nature}\ }\textbf {\bibinfo {volume} {553}},\ \bibinfo {pages} {55}
  (\bibinfo {year} {2018})}\BibitemShut {NoStop}%
\bibitem [{\citenamefont {Lohse}\ \emph {et~al.}(2015)\citenamefont {Lohse},
  \citenamefont {Schweizer}, \citenamefont {Zilberberg}, \citenamefont
  {Aidelsburger},\ and\ \citenamefont {Bloch}}]{Lohse2015}%
  \BibitemOpen
  \bibinfo {author} {M.~Lohse}, \bibinfo {author} {C.~Schweizer}, \bibinfo
  {author} {O.~Zilberberg}, \bibinfo {author} {M.~Aidelsburger}\ and\ \bibinfo
  {author} {I.~Bloch},\ \emph {\bibinfo {title} {A Thouless quantum pump with
  ultracold bosonic atoms in an optical superlattice}},\ \href
  {\doibase10.1038/nphys3584} {\bibfield  {journal} {\bibinfo  {journal}
  {Nature Physics}\ }\textbf {\bibinfo {volume} {12}},\ \bibinfo {pages} {350}
  (\bibinfo {year} {2015})}\BibitemShut {NoStop}%
\bibitem [{\citenamefont {Sato}\ \emph {et~al.}(2020)\citenamefont {Sato},
  \citenamefont {Giovannini}, \citenamefont {Aeschlimann}, \citenamefont
  {Gierz}, \citenamefont {Hübener},\ and\ \citenamefont {Rubio}}]{Sato2020}%
  \BibitemOpen
  \bibinfo {author} {S.~A. Sato}, \bibinfo {author} {U.~D. Giovannini},
  \bibinfo {author} {S.~Aeschlimann}, \bibinfo {author} {I.~Gierz}, \bibinfo
  {author} {H.~Hübener}\ and\ \bibinfo {author} {A.~Rubio},\ \emph {\bibinfo
  {title} {Floquet states in dissipative open quantum systems}},\ \href
  {\doibase10.1088/1361-6455/abb127} {\bibfield  {journal} {\bibinfo  {journal}
  {Journal of Physics B: Atomic, Molecular and Optical Physics}\ }\textbf
  {\bibinfo {volume} {53}},\ \bibinfo {pages} {225601} (\bibinfo {year}
  {2020})}\BibitemShut {NoStop}%
\bibitem [{\citenamefont {Pieplow}\ \emph {et~al.}(2019)\citenamefont
  {Pieplow}, \citenamefont {Creffield},\ and\ \citenamefont
  {Sols}}]{Pieplow2019}%
  \BibitemOpen
  \bibinfo {author} {G.~Pieplow}, \bibinfo {author} {C.~E. Creffield}\ and\
  \bibinfo {author} {F.~Sols},\ \emph {\bibinfo {title} {Protected cat states
  from kinetic driving of a boson gas}},\ \href
  {\doibase10.1103/physrevresearch.1.033013} {\bibfield  {journal} {\bibinfo
  {journal} {Physical Review Research}\ }\textbf {\bibinfo {volume} {1}},\
  \bibinfo {pages} {033013} (\bibinfo {year} {2019})}\BibitemShut {NoStop}%
\bibitem [{\citenamefont {Bai}\ \emph {et~al.}(2021)\citenamefont {Bai},
  \citenamefont {Chen}, \citenamefont {Wu},\ and\ \citenamefont
  {An}}]{Bai2021}%
  \BibitemOpen
  \bibinfo {author} {S.-Y. Bai}, \bibinfo {author} {C.~Chen}, \bibinfo {author}
  {H.~Wu}\ and\ \bibinfo {author} {J.-H. An},\ \emph {\bibinfo {title} {Quantum
  control in open and periodically driven systems}},\ \href
  {\doibase10.1080/23746149.2020.1870559} {\bibfield  {journal} {\bibinfo
  {journal} {Advances in Physics: X}\ }\textbf {\bibinfo {volume} {6}}
  (\bibinfo {year} {2021}),\ 10.1080/23746149.2020.1870559}\BibitemShut
  {NoStop}%
\bibitem [{\citenamefont {Qiu}\ \emph {et~al.}(2022)\citenamefont {Qiu},
  \citenamefont {Cheng}, \citenamefont {Chen}, \citenamefont {Lan},\ and\
  \citenamefont {Nie}}]{Qiu2022}%
  \BibitemOpen
  \bibinfo {author} {W.~Qiu}, \bibinfo {author} {X.~Cheng}, \bibinfo {author}
  {A.~Chen}, \bibinfo {author} {Y.~Lan}\ and\ \bibinfo {author} {W.~Nie},\
  \emph {\bibinfo {title} {Controlling quantum coherence and entanglement in
  cavity magnomechanical systems}},\ \href
  {\doibase10.1103/physreva.105.063718} {\bibfield  {journal} {\bibinfo
  {journal} {Physical Review A}\ }\textbf {\bibinfo {volume} {105}},\ \bibinfo
  {pages} {063718} (\bibinfo {year} {2022})}\BibitemShut {NoStop}%
\bibitem [{\citenamefont {Chen}\ \emph {et~al.}(2015)\citenamefont {Chen},
  \citenamefont {An}, \citenamefont {Luo}, \citenamefont {Sun},\ and\
  \citenamefont {Oh}}]{Chen2015}%
  \BibitemOpen
  \bibinfo {author} {C.~Chen}, \bibinfo {author} {J.-H. An}, \bibinfo {author}
  {H.-G. Luo}, \bibinfo {author} {C.~P. Sun}\ and\ \bibinfo {author} {C.~H.
  Oh},\ \emph {\bibinfo {title} {Floquet control of quantum dissipation in spin
  chains}},\ \href {\doibase10.1103/physreva.91.052122} {\bibfield  {journal}
  {\bibinfo  {journal} {Physical Review A}\ }\textbf {\bibinfo {volume} {91}},\
  \bibinfo {pages} {052122} (\bibinfo {year} {2015})}\BibitemShut {NoStop}%
\bibitem [{\citenamefont {Restrepo}\ \emph {et~al.}(2016)\citenamefont
  {Restrepo}, \citenamefont {Cerrillo}, \citenamefont {Bastidas}, \citenamefont
  {Angelakis},\ and\ \citenamefont {Brandes}}]{Restrepo2016}%
  \BibitemOpen
  \bibinfo {author} {S.~Restrepo}, \bibinfo {author} {J.~Cerrillo}, \bibinfo
  {author} {V.~Bastidas}, \bibinfo {author} {D.~Angelakis}\ and\ \bibinfo
  {author} {T.~Brandes},\ \emph {\bibinfo {title} {Driven Open Quantum Systems
  and Floquet Stroboscopic Dynamics}},\ \href
  {\doibase10.1103/physrevlett.117.250401} {\bibfield  {journal} {\bibinfo
  {journal} {Physical Review Letters}\ }\textbf {\bibinfo {volume} {117}},\
  \bibinfo {pages} {250401} (\bibinfo {year} {2016})}\BibitemShut {NoStop}%
\bibitem [{\citenamefont {Viola}\ \emph {et~al.}(1999)\citenamefont {Viola},
  \citenamefont {Knill},\ and\ \citenamefont {Lloyd}}]{Viola1999}%
  \BibitemOpen
  \bibinfo {author} {L.~Viola}, \bibinfo {author} {E.~Knill}\ and\ \bibinfo
  {author} {S.~Lloyd},\ \emph {\bibinfo {title} {Dynamical Decoupling of Open
  Quantum Systems}},\ \href {\doibase10.1103/physrevlett.82.2417} {\bibfield
  {journal} {\bibinfo  {journal} {Physical Review Letters}\ }\textbf {\bibinfo
  {volume} {82}},\ \bibinfo {pages} {2417} (\bibinfo {year}
  {1999})}\BibitemShut {NoStop}%
\bibitem [{\citenamefont {Liu}\ \emph {et~al.}(2023)\citenamefont {Liu},
  \citenamefont {Peng},\ and\ \citenamefont {Xiong}}]{Liu2023}%
  \BibitemOpen
  \bibinfo {author} {Z.-X. Liu}, \bibinfo {author} {J.~Peng}\ and\ \bibinfo
  {author} {H.~Xiong},\ \emph {\bibinfo {title} {Generation of magnonic
  frequency combs via a two-tone microwave drive}},\ \href
  {\doibase10.1103/physreva.107.053708} {\bibfield  {journal} {\bibinfo
  {journal} {Physical Review A}\ }\textbf {\bibinfo {volume} {107}},\ \bibinfo
  {pages} {053708} (\bibinfo {year} {2023})}\BibitemShut {NoStop}%
\bibitem [{\citenamefont {Wang}\ \emph {et~al.}(2020)\citenamefont {Wang},
  \citenamefont {Li},\ and\ \citenamefont {Li}}]{Wang2020}%
  \BibitemOpen
  \bibinfo {author} {X.~Wang}, \bibinfo {author} {H.-R. Li}\ and\ \bibinfo
  {author} {F.-L. Li},\ \emph {\bibinfo {title} {Generating synthetic magnetism
  via Floquet engineering auxiliary qubits in phonon-cavity-based lattice}},\
  \href {\doibase10.1088/1367-2630/ab776e} {\bibfield  {journal} {\bibinfo
  {journal} {New Journal of Physics}\ }\textbf {\bibinfo {volume} {22}},\
  \bibinfo {pages} {033037} (\bibinfo {year} {2020})}\BibitemShut {NoStop}%
\bibitem [{\citenamefont {Weitenberg}\ and\ \citenamefont
  {Simonet}(2021)}]{Weitenberg2021}%
  \BibitemOpen
  \bibinfo {author} {C.~Weitenberg}\ and\ \bibinfo {author} {J.~Simonet},\
  \emph {\bibinfo {title} {Tailoring quantum gases by {Floquet} engineering}},\
  \href {\doibase10.1038/s41567-021-01316-x} {\bibfield  {journal} {\bibinfo
  {journal} {Nature Physics}\ }\textbf {\bibinfo {volume} {17}},\ \bibinfo
  {pages} {1342} (\bibinfo {year} {2021})}\BibitemShut {NoStop}%
\bibitem [{\citenamefont {Wang}\ \emph {et~al.}(2022)\citenamefont {Wang},
  \citenamefont {Xu}, \citenamefont {Gao}, \citenamefont {Liu}, \citenamefont
  {Ma}, \citenamefont {Zhang}, \citenamefont {Wang}, \citenamefont {Cao},
  \citenamefont {Wang}, \citenamefont {Zhang}, \citenamefont {Culcer},
  \citenamefont {Hu}, \citenamefont {Jiang}, \citenamefont {Li}, \citenamefont
  {Guo},\ and\ \citenamefont {Guo}}]{Wang2022a}%
  \BibitemOpen
  \bibinfo {author} {K.~Wang}, \bibinfo {author} {G.~Xu}, \bibinfo {author}
  {F.~Gao}, \bibinfo {author} {H.~Liu}, \bibinfo {author} {R.-L. Ma}, \bibinfo
  {author} {X.~Zhang}, \bibinfo {author} {Z.~Wang}, \bibinfo {author} {G.~Cao},
  \bibinfo {author} {T.~Wang}, \bibinfo {author} {J.-J. Zhang}, \bibinfo
  {author} {D.~Culcer}, \bibinfo {author} {X.~Hu}, \bibinfo {author} {H.-W.
  Jiang}, \bibinfo {author} {H.-O. Li}, \bibinfo {author} {G.-C. Guo}\ and\
  \bibinfo {author} {G.-P. Guo},\ \emph {\bibinfo {title} {Ultrafast coherent
  control of a hole spin qubit in a germanium quantum dot}},\ \href
  {\doibase10.1038/s41467-021-27880-7} {\bibfield  {journal} {\bibinfo
  {journal} {Nature Communications}\ }\textbf {\bibinfo {volume} {13}},\
  \bibinfo {pages} {206} (\bibinfo {year} {2022})}\BibitemShut {NoStop}%
\bibitem [{\citenamefont {Eichler}\ \emph {et~al.}(2018)\citenamefont
  {Eichler}, \citenamefont {Heugel}, \citenamefont {Leuch}, \citenamefont
  {Degen}, \citenamefont {Chitra},\ and\ \citenamefont
  {Zilberberg}}]{eichler2018}%
  \BibitemOpen
  \bibinfo {author} {A.~Eichler}, \bibinfo {author} {T.~L. Heugel}, \bibinfo
  {author} {A.~Leuch}, \bibinfo {author} {C.~L. Degen}, \bibinfo {author}
  {R.~Chitra}\ and\ \bibinfo {author} {O.~Zilberberg},\ \emph {\bibinfo {title}
  {A parametric symmetry breaking transducer}},\ \href
  {\doibase10.1063/1.5031058} {\bibfield  {journal} {\bibinfo  {journal}
  {Applied Physics Letters}\ }\textbf {\bibinfo {volume} {112}},\ \bibinfo
  {pages} {233105} (\bibinfo {year} {2018})}\BibitemShut {NoStop}%
\bibitem [{\citenamefont {Heugel}\ \emph {et~al.}(2019)\citenamefont {Heugel},
  \citenamefont {Biondi}, \citenamefont {Zilberberg},\ and\ \citenamefont
  {Chitra}}]{heugel2019quantum}%
  \BibitemOpen
  \bibinfo {author} {T.~L. Heugel}, \bibinfo {author} {M.~Biondi}, \bibinfo
  {author} {O.~Zilberberg}\ and\ \bibinfo {author} {R.~Chitra},\ \emph
  {\bibinfo {title} {Quantum transducer using a parametric driven-dissipative
  phase transition}},\ \href@noop {} {\bibfield  {journal} {\bibinfo  {journal}
  {Physical review letters}\ }\textbf {\bibinfo {volume} {123}},\ \bibinfo
  {pages} {173601} (\bibinfo {year} {2019})}\BibitemShut {NoStop}%
\bibitem [{\citenamefont {Bai}\ and\ \citenamefont {An}(2023)}]{Bai2023}%
  \BibitemOpen
  \bibinfo {author} {S.-Y. Bai}\ and\ \bibinfo {author} {J.-H. An},\ \href
  {\doibase10.48550/ARXIV.2303.00392} {\emph {\bibinfo {title} {Floquet
  engineering to overcome no-go theorem of noisy quantum metrology}}} (\bibinfo
  {year} {2023})\BibitemShut {NoStop}%
\bibitem [{\citenamefont {Rudner}\ and\ \citenamefont
  {Lindner}(2020{\natexlab{b}})}]{Rudner2020a}%
  \BibitemOpen
  \bibinfo {author} {M.~S. Rudner}\ and\ \bibinfo {author} {N.~H. Lindner},\
  \emph {\bibinfo {title} {Band structure engineering and non-equilibrium
  dynamics in {Floquet} topological insulators}},\ \href
  {\doibase10.1038/s42254-020-0170-z} {\bibfield  {journal} {\bibinfo
  {journal} {Nature Reviews Physics}\ }\textbf {\bibinfo {volume} {2}},\
  \bibinfo {pages} {229} (\bibinfo {year} {2020}{\natexlab{b}})}\BibitemShut
  {NoStop}%
\bibitem [{\citenamefont {Harper}\ \emph {et~al.}(2020)\citenamefont {Harper},
  \citenamefont {Roy}, \citenamefont {Rudner},\ and\ \citenamefont
  {Sondhi}}]{Harper2020}%
  \BibitemOpen
  \bibinfo {author} {F.~Harper}, \bibinfo {author} {R.~Roy}, \bibinfo {author}
  {M.~S. Rudner}\ and\ \bibinfo {author} {S.~Sondhi},\ \emph {\bibinfo {title}
  {Topology and Broken Symmetry in Floquet Systems}},\ \href
  {\doibase10.1146/annurev-conmatphys-031218-013721} {\bibfield  {journal}
  {\bibinfo  {journal} {Annual Review of Condensed Matter Physics}\ }\textbf
  {\bibinfo {volume} {11}},\ \bibinfo {pages} {345} (\bibinfo {year}
  {2020})}\BibitemShut {NoStop}%
\bibitem [{\citenamefont {Kitagawa}\ \emph {et~al.}(2010)\citenamefont
  {Kitagawa}, \citenamefont {Berg}, \citenamefont {Rudner},\ and\ \citenamefont
  {Demler}}]{Kitagawa2010}%
  \BibitemOpen
  \bibinfo {author} {T.~Kitagawa}, \bibinfo {author} {E.~Berg}, \bibinfo
  {author} {M.~Rudner}\ and\ \bibinfo {author} {E.~Demler},\ \emph {\bibinfo
  {title} {Topological characterization of periodically driven quantum
  systems}},\ \href {\doibase10.1103/physrevb.82.235114} {\bibfield  {journal}
  {\bibinfo  {journal} {Physical Review B}\ }\textbf {\bibinfo {volume} {82}},\
  \bibinfo {pages} {235114} (\bibinfo {year} {2010})}\BibitemShut {NoStop}%
\bibitem [{\citenamefont {Lindner}\ \emph {et~al.}(2011)\citenamefont
  {Lindner}, \citenamefont {Refael},\ and\ \citenamefont
  {Galitski}}]{Lindner2011}%
  \BibitemOpen
  \bibinfo {author} {N.~H. Lindner}, \bibinfo {author} {G.~Refael}\ and\
  \bibinfo {author} {V.~Galitski},\ \emph {\bibinfo {title} {Floquet
  topological insulator in semiconductor quantum wells}},\ \href
  {\doibase10.1038/nphys1926} {\bibfield  {journal} {\bibinfo  {journal}
  {Nature Physics}\ }\textbf {\bibinfo {volume} {7}},\ \bibinfo {pages} {490}
  (\bibinfo {year} {2011})}\BibitemShut {NoStop}%
\bibitem [{\citenamefont {Del~Pino}\ and\ \citenamefont
  {Zilberberg}(2023)}]{del2023dynamical}%
  \BibitemOpen
  \bibinfo {author} {J.~Del~Pino}\ and\ \bibinfo {author} {O.~Zilberberg},\
  \emph {\bibinfo {title} {Dynamical gauge fields with bosonic codes}},\
  \href@noop {} {\bibfield  {journal} {\bibinfo  {journal} {Physical Review
  Letters}\ }\textbf {\bibinfo {volume} {130}},\ \bibinfo {pages} {171901}
  (\bibinfo {year} {2023})}\BibitemShut {NoStop}%
\bibitem [{\citenamefont {Grimm}\ \emph {et~al.}(2020)\citenamefont {Grimm},
  \citenamefont {Frattini}, \citenamefont {Puri}, \citenamefont {Mundhada},
  \citenamefont {Touzard}, \citenamefont {Mirrahimi}, \citenamefont {Girvin},
  \citenamefont {Shankar},\ and\ \citenamefont {Devoret}}]{Grimm2020}%
  \BibitemOpen
  \bibinfo {author} {A.~Grimm}, \bibinfo {author} {N.~E. Frattini}, \bibinfo
  {author} {S.~Puri}, \bibinfo {author} {S.~O. Mundhada}, \bibinfo {author}
  {S.~Touzard}, \bibinfo {author} {M.~Mirrahimi}, \bibinfo {author} {S.~M.
  Girvin}, \bibinfo {author} {S.~Shankar}\ and\ \bibinfo {author} {M.~H.
  Devoret},\ \emph {\bibinfo {title} {Stabilization and operation of a Kerr-cat
  qubit}},\ \href {\doibase10.1038/s41586-020-2587-z} {\bibfield  {journal}
  {\bibinfo  {journal} {Nature}\ }\textbf {\bibinfo {volume} {584}},\ \bibinfo
  {pages} {205} (\bibinfo {year} {2020})}\BibitemShut {NoStop}%
\bibitem [{\citenamefont {Mattes}\ \emph {et~al.}(2024)\citenamefont {Mattes},
  \citenamefont {Volkov},\ and\ \citenamefont {Baum}}]{Mattes2024}%
  \BibitemOpen
  \bibinfo {author} {M.~Mattes}, \bibinfo {author} {M.~Volkov}\ and\ \bibinfo
  {author} {P.~Baum},\ \emph {\bibinfo {title} {Femtosecond electron beam probe
  of ultrafast electronics}},\ \href {\doibase10.1038/s41467-024-45744-8}
  {\bibfield  {journal} {\bibinfo  {journal} {Nature Communications}\ }\textbf
  {\bibinfo {volume} {15}},\ \bibinfo {pages} {1743} (\bibinfo {year}
  {2024})}\BibitemShut {NoStop}%
\bibitem [{\citenamefont {Budden}\ \emph {et~al.}(2021)\citenamefont {Budden},
  \citenamefont {Gebert}, \citenamefont {Buzzi}, \citenamefont {Jotzu},
  \citenamefont {Wang}, \citenamefont {Matsuyama}, \citenamefont {Meier},
  \citenamefont {Laplace}, \citenamefont {Pontiroli}, \citenamefont {Riccò},
  \citenamefont {Schlawin}, \citenamefont {Jaksch},\ and\ \citenamefont
  {Cavalleri}}]{Budden2021}%
  \BibitemOpen
  \bibinfo {author} {M.~Budden}, \bibinfo {author} {T.~Gebert}, \bibinfo
  {author} {M.~Buzzi}, \bibinfo {author} {G.~Jotzu}, \bibinfo {author}
  {E.~Wang}, \bibinfo {author} {T.~Matsuyama}, \bibinfo {author} {G.~Meier},
  \bibinfo {author} {Y.~Laplace}, \bibinfo {author} {D.~Pontiroli}, \bibinfo
  {author} {M.~Riccò}, \bibinfo {author} {F.~Schlawin}, \bibinfo {author}
  {D.~Jaksch}\ and\ \bibinfo {author} {A.~Cavalleri},\ \emph {\bibinfo {title}
  {Evidence for metastable photo-induced superconductivity in {K3C60}}},\ \href
  {\doibase10.1038/s41567-020-01148-1} {\bibfield  {journal} {\bibinfo
  {journal} {Nature Physics}\ }\textbf {\bibinfo {volume} {17}},\ \bibinfo
  {pages} {611} (\bibinfo {year} {2021})}\BibitemShut {NoStop}%
\bibitem [{\citenamefont {Bloch}\ and\ \citenamefont
  {Siegert}(1940)}]{BlochSiegert1940}%
  \BibitemOpen
  \bibinfo {author} {F.~Bloch}\ and\ \bibinfo {author} {A.~Siegert},\ \emph
  {\bibinfo {title} {Magnetic {Resonance} for {Nonrotating} {Fields}}},\ \href
  {\doibase10.1103/PhysRev.57.522} {\bibfield  {journal} {\bibinfo  {journal}
  {Physical Review}\ }\textbf {\bibinfo {volume} {57}},\ \bibinfo {pages} {522}
  (\bibinfo {year} {1940})}\BibitemShut {NoStop}%
\bibitem [{\citenamefont {Eckardt}\ and\ \citenamefont
  {Anisimovas}(2015)}]{Eckardt2015}%
  \BibitemOpen
  \bibinfo {author} {A.~Eckardt}\ and\ \bibinfo {author} {E.~Anisimovas},\
  \emph {\bibinfo {title} {High-frequency approximation for periodically driven
  quantum systems from a {Floquet}-space perspective}},\ \href
  {\doibase10.1088/1367-2630/17/9/093039} {\bibfield  {journal} {\bibinfo
  {journal} {New Journal of Physics}\ }\textbf {\bibinfo {volume} {17}},\
  \bibinfo {pages} {093039} (\bibinfo {year} {2015})}\BibitemShut {NoStop}%
\bibitem [{\citenamefont {Casas}\ \emph {et~al.}(2001)\citenamefont {Casas},
  \citenamefont {Oteo},\ and\ \citenamefont {Ros}}]{Casas2001}%
  \BibitemOpen
  \bibinfo {author} {F.~Casas}, \bibinfo {author} {J.~A. Oteo}\ and\ \bibinfo
  {author} {J.~Ros},\ \emph {\bibinfo {title} {Floquet theory: exponential
  perturbative treatment}},\ \href {\doibase10.1088/0305-4470/34/16/305}
  {\bibfield  {journal} {\bibinfo  {journal} {Journal of Physics A:
  Mathematical and General}\ }\textbf {\bibinfo {volume} {34}},\ \bibinfo
  {pages} {3379} (\bibinfo {year} {2001})}\BibitemShut {NoStop}%
\bibitem [{\citenamefont {Blanes}\ \emph {et~al.}(2009)\citenamefont {Blanes},
  \citenamefont {Casas}, \citenamefont {Oteo},\ and\ \citenamefont
  {Ros}}]{Blanes2009}%
  \BibitemOpen
  \bibinfo {author} {S.~Blanes}, \bibinfo {author} {F.~Casas}, \bibinfo
  {author} {J.~A. Oteo}\ and\ \bibinfo {author} {J.~Ros},\ \emph {\bibinfo
  {title} {The {Magnus} expansion and some of its applications}},\ \href
  {\doibase10.1016/j.physrep.2008.11.001} {\bibfield  {journal} {\bibinfo
  {journal} {Physics Reports}\ }\textbf {\bibinfo {volume} {470}},\ \bibinfo
  {pages} {151} (\bibinfo {year} {2009})}\BibitemShut {NoStop}%
\bibitem [{\citenamefont {Mikami}\ \emph {et~al.}(2016)\citenamefont {Mikami},
  \citenamefont {Kitamura}, \citenamefont {Yasuda}, \citenamefont {Tsuji},
  \citenamefont {Oka},\ and\ \citenamefont {Aoki}}]{Mikami2016}%
  \BibitemOpen
  \bibinfo {author} {T.~Mikami}, \bibinfo {author} {S.~Kitamura}, \bibinfo
  {author} {K.~Yasuda}, \bibinfo {author} {N.~Tsuji}, \bibinfo {author}
  {T.~Oka}\ and\ \bibinfo {author} {H.~Aoki},\ \emph {\bibinfo {title}
  {Brillouin-Wigner theory for high-frequency expansion in periodically driven
  systems: Application to Floquet topological insulators}},\ \href
  {\doibase10.1103/physrevb.93.144307} {\bibfield  {journal} {\bibinfo
  {journal} {Physical Review B}\ }\textbf {\bibinfo {volume} {93}},\ \bibinfo
  {pages} {144307} (\bibinfo {year} {2016})}\BibitemShut {NoStop}%
\bibitem [{\citenamefont {Bukov}\ \emph {et~al.}(2015)\citenamefont {Bukov},
  \citenamefont {D'Alessio},\ and\ \citenamefont {Polkovnikov}}]{Bukov2015}%
  \BibitemOpen
  \bibinfo {author} {M.~Bukov}, \bibinfo {author} {L.~D'Alessio}\ and\ \bibinfo
  {author} {A.~Polkovnikov},\ \emph {\bibinfo {title} {Universal high-frequency
  behavior of periodically driven systems: from dynamical stabilization to
  {Floquet} engineering}},\ \href {\doibase10.1080/00018732.2015.1055918}
  {\bibfield  {journal} {\bibinfo  {journal} {Advances in Physics}\ }\textbf
  {\bibinfo {volume} {64}},\ \bibinfo {pages} {139} (\bibinfo {year}
  {2015})}\BibitemShut {NoStop}%
\bibitem [{\citenamefont {Goldman}\ and\ \citenamefont
  {Dalibard}(2014)}]{Goldman2014}%
  \BibitemOpen
  \bibinfo {author} {N.~Goldman}\ and\ \bibinfo {author} {J.~Dalibard},\ \emph
  {\bibinfo {title} {Periodically Driven Quantum Systems: Effective
  Hamiltonians and Engineered Gauge Fields}},\ \href
  {\doibase10.1103/physrevx.4.031027} {\bibfield  {journal} {\bibinfo
  {journal} {Physical Review X}\ }\textbf {\bibinfo {volume} {4}},\ \bibinfo
  {pages} {031027} (\bibinfo {year} {2014})}\BibitemShut {NoStop}%
\bibitem [{\citenamefont {moo Ann}\ \emph {et~al.}(2021)\citenamefont {moo
  Ann}, \citenamefont {Kessels},\ and\ \citenamefont {Steele}}]{Ann2021}%
  \BibitemOpen
  \bibinfo {author} {B.~moo Ann}, \bibinfo {author} {W.~Kessels}\ and\ \bibinfo
  {author} {G.~A. Steele},\ \emph {\bibinfo {title} {Sideband transitions in a
  two-mode Josephson circuit driven beyond the rotating-wave approximation}},\
  \href {\doibase10.1103/physrevresearch.3.033004} {\bibfield  {journal}
  {\bibinfo  {journal} {Physical Review Research}\ }\textbf {\bibinfo {volume}
  {3}},\ \bibinfo {pages} {033004} (\bibinfo {year} {2021})}\BibitemShut
  {NoStop}%
\bibitem [{\citenamefont {Zilberberg}\ and\ \citenamefont
  {Eichler}(2023)}]{Zilberberg2023}%
  \BibitemOpen
  \bibinfo {author} {O.~Zilberberg}\ and\ \bibinfo {author} {A.~Eichler},\
  \href@noop {} {\emph {\bibinfo {title} {Classical and {Quantum} {Parametric}
  {Phenomena}}}},\ Oxford {Graduate} {Texts}\ (\bibinfo  {publisher} {Oxford
  University Press},\ \bibinfo {address} {Oxford, New York},\ \bibinfo {year}
  {2023})\BibitemShut {NoStop}%
\bibitem [{\citenamefont {Krylov}\ and\ \citenamefont
  {Bogoliubov}(1949)}]{Krylov1949}%
  \BibitemOpen
  \bibinfo {author} {N.~M. Krylov}\ and\ \bibinfo {author} {N.~N. Bogoliubov},\
  \href@noop {} {\emph {\bibinfo {title} {Introduction to Non-Linear Mechanics.
  (AM-11) (Annals of Mathematics Studies)}}}\ (\bibinfo  {publisher} {Princeton
  University Press},\ \bibinfo {year} {1949})\ p.\ \bibinfo {pages}
  {106}\BibitemShut {NoStop}%
\bibitem [{\citenamefont {Bogoliubov}\ and\ \citenamefont
  {Mitropol'skii}(1961)}]{Bogoliubov1961}%
  \BibitemOpen
  \bibinfo {author} {N.~N. Bogoliubov}\ and\ \bibinfo {author} {I.~A.
  Mitropol'skii},\ \href@noop {} {{\selectlanguage {English}\emph {\bibinfo
  {title} {Asymptotic {Methods} in the {Theory} of {Nonlinear}
  {Oscillations}}}}}\ (\bibinfo  {publisher} {Gordon \& Breach Science
  Publishers Ltd},\ \bibinfo {year} {1961})\BibitemShut {NoStop}%
\bibitem [{\citenamefont {Sanders}\ \emph {et~al.}(2007)\citenamefont
  {Sanders}, \citenamefont {Verhulst},\ and\ \citenamefont
  {Murdock}}]{Sanders2007}%
  \BibitemOpen
  \bibinfo {author} {J.~A. Sanders}, \bibinfo {author} {F.~Verhulst}\ and\
  \bibinfo {author} {J.~Murdock},\ \href@noop {} {\emph {\bibinfo {title}
  {Averaging Methods in Nonlinear Dynamical Systems (Applied Mathematical
  Sciences)}}}\ (\bibinfo  {publisher} {Springer},\ \bibinfo {year} {2007})\
  p.\ \bibinfo {pages} {434}\BibitemShut {NoStop}%
\bibitem [{\citenamefont {Guckenheimer}\ and\ \citenamefont
  {Holmes}(2013)}]{Guckenheimer2013}%
  \BibitemOpen
  \bibinfo {author} {J.~Guckenheimer}\ and\ \bibinfo {author} {P.~Holmes},\
  \href@noop {} {\emph {\bibinfo {title} {Nonlinear Oscillations, Dynamical
  Systems, and Bifurcations of Vector Fields}}}\ (\bibinfo  {publisher}
  {Springer London, Limited},\ \bibinfo {year} {2013})\BibitemShut {NoStop}%
\bibitem [{\citenamefont {Oliveira}(2017)}]{Oliveira2017}%
  \BibitemOpen
  \bibinfo {author} {A.~R.~E. Oliveira},\ \emph {\bibinfo {title} {History of
  Krylov-Bogoliubov-Mitropolsky Methods of Nonlinear Oscillations}},\ \href
  {\doibase10.4236/ahs.2017.61003} {\bibfield  {journal} {\bibinfo  {journal}
  {Advances in Historical Studies}\ }\textbf {\bibinfo {volume} {06}},\
  \bibinfo {pages} {40} (\bibinfo {year} {2017})}\BibitemShut {NoStop}%
\bibitem [{\citenamefont {Neu}(1980)}]{Neu1980}%
  \BibitemOpen
  \bibinfo {author} {J.~C. Neu},\ \emph {\bibinfo {title} {The Method of
  Near-Identity Transformations and Its Applications}},\ \href
  {\doibase10.1137/0138017} {\bibfield  {journal} {\bibinfo  {journal} {{SIAM}
  Journal on Applied Mathematics}\ }\textbf {\bibinfo {volume} {38}},\ \bibinfo
  {pages} {189} (\bibinfo {year} {1980})}\BibitemShut {NoStop}%
\bibitem [{\citenamefont {Holmes}\ and\ \citenamefont
  {Holmes}(1981)}]{Holmes1981}%
  \BibitemOpen
  \bibinfo {author} {C.~Holmes}\ and\ \bibinfo {author} {P.~Holmes},\ \emph
  {\bibinfo {title} {Second order averaging and bifurcations to subharmonics in
  duffing{\textquotesingle}s equation}},\ \href
  {\doibase10.1016/s0022-460x(81)80030-2} {\bibfield  {journal} {\bibinfo
  {journal} {Journal of Sound and Vibration}\ }\textbf {\bibinfo {volume}
  {78}},\ \bibinfo {pages} {161} (\bibinfo {year} {1981})}\BibitemShut
  {NoStop}%
\bibitem [{\citenamefont {Ford}\ and\ \citenamefont
  {O{\textquotesingle}Connell}(1996)}]{Ford1996}%
  \BibitemOpen
  \bibinfo {author} {G.~Ford}\ and\ \bibinfo {author}
  {R.~O{\textquotesingle}Connell},\ \emph {\bibinfo {title} {Inconsistency of
  the rotating wave approximation with the Ehrenfest theorem}},\ \href
  {\doibase10.1016/0375-9601(96)00242-3} {\bibfield  {journal} {\bibinfo
  {journal} {Physics Letters A}\ }\textbf {\bibinfo {volume} {215}},\ \bibinfo
  {pages} {245} (\bibinfo {year} {1996})}\BibitemShut {NoStop}%
\bibitem [{\citenamefont {J{\o}rgensen}\ and\ \citenamefont
  {Wubs}(2022)}]{Joergensen2022}%
  \BibitemOpen
  \bibinfo {author} {M.~A. J{\o}rgensen}\ and\ \bibinfo {author} {M.~Wubs},\
  \emph {\bibinfo {title} {Quantifying the breakdown of the rotating-wave
  approximation in single-photon superradiance}},\ \href
  {\doibase10.1088/1361-6455/ac89be} {\bibfield  {journal} {\bibinfo  {journal}
  {Journal of Physics B: Atomic, Molecular and Optical Physics}\ }\textbf
  {\bibinfo {volume} {55}},\ \bibinfo {pages} {195401} (\bibinfo {year}
  {2022})}\BibitemShut {NoStop}%
\bibitem [{\citenamefont {Laucht}\ \emph {et~al.}(2016)\citenamefont {Laucht},
  \citenamefont {Simmons}, \citenamefont {Kalra}, \citenamefont {Tosi},
  \citenamefont {Dehollain}, \citenamefont {Muhonen}, \citenamefont {Freer},
  \citenamefont {Hudson}, \citenamefont {Itoh}, \citenamefont {Jamieson},
  \citenamefont {McCallum}, \citenamefont {Dzurak},\ and\ \citenamefont
  {Morello}}]{Laucht2016}%
  \BibitemOpen
  \bibinfo {author} {A.~Laucht}, \bibinfo {author} {S.~Simmons}, \bibinfo
  {author} {R.~Kalra}, \bibinfo {author} {G.~Tosi}, \bibinfo {author} {J.~P.
  Dehollain}, \bibinfo {author} {J.~T. Muhonen}, \bibinfo {author} {S.~Freer},
  \bibinfo {author} {F.~E. Hudson}, \bibinfo {author} {K.~M. Itoh}, \bibinfo
  {author} {D.~N. Jamieson}, \bibinfo {author} {J.~C. McCallum}, \bibinfo
  {author} {A.~S. Dzurak}\ and\ \bibinfo {author} {A.~Morello},\ \emph
  {\bibinfo {title} {Breaking the rotating wave approximation for a strongly
  driven dressed single-electron spin}},\ \href
  {\doibase10.1103/physrevb.94.161302} {\bibfield  {journal} {\bibinfo
  {journal} {Physical Review B}\ }\textbf {\bibinfo {volume} {94}},\ \bibinfo
  {pages} {161302} (\bibinfo {year} {2016})}\BibitemShut {NoStop}%
\bibitem [{\citenamefont {Sharaby}\ \emph {et~al.}(2010)\citenamefont
  {Sharaby}, \citenamefont {Joshi},\ and\ \citenamefont
  {Hassan}}]{Sharaby2010}%
  \BibitemOpen
  \bibinfo {author} {Y.~A. Sharaby}, \bibinfo {author} {A.~Joshi}\ and\
  \bibinfo {author} {S.~S. Hassan},\ \emph {\bibinfo {title} {Optical
  bistability without the rotating wave approximation}},\ \href
  {\doibase10.1016/j.physleta.2010.03.022} {\bibfield  {journal} {\bibinfo
  {journal} {Physics Letters A}\ }\textbf {\bibinfo {volume} {374}},\ \bibinfo
  {pages} {2188} (\bibinfo {year} {2010})}\BibitemShut {NoStop}%
\bibitem [{\citenamefont {Zheng}\ \emph {et~al.}(2008)\citenamefont {Zheng},
  \citenamefont {Zhu},\ and\ \citenamefont {Zubairy}}]{Zheng2008}%
  \BibitemOpen
  \bibinfo {author} {H.~Zheng}, \bibinfo {author} {S.~Y. Zhu}\ and\ \bibinfo
  {author} {M.~S. Zubairy},\ \emph {\bibinfo {title} {Quantum Zeno and
  Anti-Zeno Effects: Without the Rotating-Wave Approximation}},\ \href
  {\doibase10.1103/physrevlett.101.200404} {\bibfield  {journal} {\bibinfo
  {journal} {Physical Review Letters}\ }\textbf {\bibinfo {volume} {101}},\
  \bibinfo {pages} {200404} (\bibinfo {year} {2008})}\BibitemShut {NoStop}%
\bibitem [{\citenamefont {Zhang}\ and\ \citenamefont {Chen}(2015)}]{Zhang2015}%
  \BibitemOpen
  \bibinfo {author} {Y.-Y. Zhang}\ and\ \bibinfo {author} {Q.-H. Chen},\ \emph
  {\bibinfo {title} {Generalized rotating-wave approximation for the two-qubit
  quantum Rabi model}},\ \href {\doibase10.1103/physreva.91.013814} {\bibfield
  {journal} {\bibinfo  {journal} {Physical Review A}\ }\textbf {\bibinfo
  {volume} {91}},\ \bibinfo {pages} {013814} (\bibinfo {year}
  {2015})}\BibitemShut {NoStop}%
\bibitem [{\citenamefont {Zueco}\ \emph {et~al.}(2009)\citenamefont {Zueco},
  \citenamefont {Reuther}, \citenamefont {Kohler},\ and\ \citenamefont
  {Hänggi}}]{Zueco2009}%
  \BibitemOpen
  \bibinfo {author} {D.~Zueco}, \bibinfo {author} {G.~M. Reuther}, \bibinfo
  {author} {S.~Kohler}\ and\ \bibinfo {author} {P.~Hänggi},\ \emph {\bibinfo
  {title} {Qubit-oscillator dynamics in the dispersive regime: Analytical
  theory beyond the rotating-wave approximation}},\ \href
  {\doibase10.1103/physreva.80.033846} {\bibfield  {journal} {\bibinfo
  {journal} {Physical Review A}\ }\textbf {\bibinfo {volume} {80}},\ \bibinfo
  {pages} {033846} (\bibinfo {year} {2009})}\BibitemShut {NoStop}%
\bibitem [{\citenamefont {Zeuch}\ \emph {et~al.}(2020)\citenamefont {Zeuch},
  \citenamefont {Hassler}, \citenamefont {Slim},\ and\ \citenamefont
  {DiVincenzo}}]{Zeuch2020}%
  \BibitemOpen
  \bibinfo {author} {D.~Zeuch}, \bibinfo {author} {F.~Hassler}, \bibinfo
  {author} {J.~J. Slim}\ and\ \bibinfo {author} {D.~P. DiVincenzo},\ \emph
  {\bibinfo {title} {Exact rotating wave approximation}},\ \href
  {\doibase10.1016/j.aop.2020.168327} {\bibfield  {journal} {\bibinfo
  {journal} {Annals of Physics}\ }\textbf {\bibinfo {volume} {423}},\ \bibinfo
  {pages} {168327} (\bibinfo {year} {2020})}\BibitemShut {NoStop}%
\bibitem [{\citenamefont {Wang}\ \emph {et~al.}(2021)\citenamefont {Wang},
  \citenamefont {Yin}, \citenamefont {Yang}, \citenamefont {Ji},\ and\
  \citenamefont {Sun}}]{Wang2021}%
  \BibitemOpen
  \bibinfo {author} {Y.-F. Wang}, \bibinfo {author} {H.-H. Yin}, \bibinfo
  {author} {M.-Y. Yang}, \bibinfo {author} {A.-C. Ji}\ and\ \bibinfo {author}
  {Q.~Sun},\ \emph {\bibinfo {title} {Effective Hamiltonian of the
  Jaynes-Cummings model beyond rotating-wave approximation}},\ \href
  {\doibase10.1088/1674-1056/abd930} {\bibfield  {journal} {\bibinfo  {journal}
  {Chinese Physics B}\ }\textbf {\bibinfo {volume} {30}},\ \bibinfo {pages}
  {064204} (\bibinfo {year} {2021})}\BibitemShut {NoStop}%
\bibitem [{\citenamefont {Gan}\ and\ \citenamefont {Zheng}(2010)}]{Gan2010}%
  \BibitemOpen
  \bibinfo {author} {C.~J. Gan}\ and\ \bibinfo {author} {H.~Zheng},\ \emph
  {\bibinfo {title} {Dynamics of a two-level system coupled to a quantum
  oscillator: transformed rotating-wave approximation}},\ \href
  {\doibase10.1140/epjd/e2010-00182-8} {\bibfield  {journal} {\bibinfo
  {journal} {The European Physical Journal D}\ }\textbf {\bibinfo {volume}
  {59}},\ \bibinfo {pages} {473} (\bibinfo {year} {2010})}\BibitemShut
  {NoStop}%
\bibitem [{\citenamefont {He}\ \emph {et~al.}(2012)\citenamefont {He},
  \citenamefont {Wang}, \citenamefont {Chen}, \citenamefont {Ren},
  \citenamefont {Liu},\ and\ \citenamefont {Wang}}]{He2012}%
  \BibitemOpen
  \bibinfo {author} {S.~He}, \bibinfo {author} {C.~Wang}, \bibinfo {author}
  {Q.-H. Chen}, \bibinfo {author} {X.-Z. Ren}, \bibinfo {author} {T.~Liu}\ and\
  \bibinfo {author} {K.-L. Wang},\ \emph {\bibinfo {title} {First-order
  corrections to the rotating-wave approximation in the Jaynes-Cummings
  model}},\ \href {\doibase10.1103/physreva.86.033837} {\bibfield  {journal}
  {\bibinfo  {journal} {Physical Review A}\ }\textbf {\bibinfo {volume} {86}},\
  \bibinfo {pages} {033837} (\bibinfo {year} {2012})}\BibitemShut {NoStop}%
\bibitem [{\citenamefont {Hausinger}\ and\ \citenamefont
  {Grifoni}(2008)}]{Hausinger2008}%
  \BibitemOpen
  \bibinfo {author} {J.~Hausinger}\ and\ \bibinfo {author} {M.~Grifoni},\ \emph
  {\bibinfo {title} {Dissipative dynamics of a biased qubit coupled to a
  harmonic oscillator: analytical results beyond the rotating wave
  approximation}},\ \href {\doibase10.1088/1367-2630/10/11/115015} {\bibfield
  {journal} {\bibinfo  {journal} {New Journal of Physics}\ }\textbf {\bibinfo
  {volume} {10}},\ \bibinfo {pages} {115015} (\bibinfo {year}
  {2008})}\BibitemShut {NoStop}%
\bibitem [{\citenamefont {Nourmandipour}\ \emph {et~al.}(2021)\citenamefont
  {Nourmandipour}, \citenamefont {Vafafard}, \citenamefont {Mortezapour},\ and\
  \citenamefont {Franzosi}}]{Nourmandipour2021}%
  \BibitemOpen
  \bibinfo {author} {A.~Nourmandipour}, \bibinfo {author} {A.~Vafafard},
  \bibinfo {author} {A.~Mortezapour}\ and\ \bibinfo {author} {R.~Franzosi},\
  \emph {\bibinfo {title} {Entanglement protection of classically driven qubits
  in a lossy cavity}},\ \href {\doibase10.1038/s41598-021-95623-1} {\bibfield
  {journal} {\bibinfo  {journal} {Scientific Reports}\ }\textbf {\bibinfo
  {volume} {11}},\ \bibinfo {pages} {16259} (\bibinfo {year}
  {2021})}\BibitemShut {NoStop}%
\bibitem [{\citenamefont {Cortiñas}(2024)}]{Cortinas2024}%
  \BibitemOpen
  \bibinfo {author} {R.~G. Cortiñas},\ \emph {\bibinfo {title} {Towards the
  generation of mechanical {Kerr}-cats: awakening the perturbative quantum
  {Moyal} corrections to classical motion}},\ \href
  {\doibase10.1088/1367-2630/ad1e90} {\bibfield  {journal} {\bibinfo  {journal}
  {New Journal of Physics}\ }\textbf {\bibinfo {volume} {26}},\ \bibinfo
  {pages} {023022} (\bibinfo {year} {2024})}\BibitemShut {NoStop}%
\bibitem [{\citenamefont {Zhang}\ \emph {et~al.}(2021)\citenamefont {Zhang},
  \citenamefont {Qin}, \citenamefont {Song},\ and\ \citenamefont
  {Long}}]{Zhang2021}%
  \BibitemOpen
  \bibinfo {author} {H.~Zhang}, \bibinfo {author} {G.-Q. Qin}, \bibinfo
  {author} {X.-K. Song}\ and\ \bibinfo {author} {G.-L. Long},\ \emph {\bibinfo
  {title} {Color-detuning-dynamics-based quantum sensing with dressed states
  driving}},\ \href {\doibase10.1364/OE.413637} {\bibfield  {journal} {\bibinfo
   {journal} {Optics Express}\ }\textbf {\bibinfo {volume} {29}},\ \bibinfo
  {pages} {5358} (\bibinfo {year} {2021})}\BibitemShut {NoStop}%
\bibitem [{\citenamefont {Schindler}\ \emph {et~al.}(2013)\citenamefont
  {Schindler}, \citenamefont {Müller}, \citenamefont {Nigg}, \citenamefont
  {Barreiro}, \citenamefont {Martinez}, \citenamefont {Hennrich}, \citenamefont
  {Monz}, \citenamefont {Diehl}, \citenamefont {Zoller},\ and\ \citenamefont
  {Blatt}}]{Schindler2013}%
  \BibitemOpen
  \bibinfo {author} {P.~Schindler}, \bibinfo {author} {M.~Müller}, \bibinfo
  {author} {D.~Nigg}, \bibinfo {author} {J.~T. Barreiro}, \bibinfo {author}
  {E.~A. Martinez}, \bibinfo {author} {M.~Hennrich}, \bibinfo {author}
  {T.~Monz}, \bibinfo {author} {S.~Diehl}, \bibinfo {author} {P.~Zoller}\ and\
  \bibinfo {author} {R.~Blatt},\ \emph {\bibinfo {title} {Quantum simulation of
  dynamical maps with trapped ions}},\ \href {\doibase10.1038/nphys2630}
  {\bibfield  {journal} {\bibinfo  {journal} {Nature Physics}\ }\textbf
  {\bibinfo {volume} {9}},\ \bibinfo {pages} {361} (\bibinfo {year}
  {2013})}\BibitemShut {NoStop}%
\bibitem [{\citenamefont {Sandholzer}\ \emph {et~al.}(2019)\citenamefont
  {Sandholzer}, \citenamefont {Murakami}, \citenamefont {Görg}, \citenamefont
  {Minguzzi}, \citenamefont {Messer}, \citenamefont {Desbuquois}, \citenamefont
  {Eckstein}, \citenamefont {Werner},\ and\ \citenamefont
  {Esslinger}}]{Sandholzer2019}%
  \BibitemOpen
  \bibinfo {author} {K.~Sandholzer}, \bibinfo {author} {Y.~Murakami}, \bibinfo
  {author} {F.~Görg}, \bibinfo {author} {J.~Minguzzi}, \bibinfo {author}
  {M.~Messer}, \bibinfo {author} {R.~Desbuquois}, \bibinfo {author}
  {M.~Eckstein}, \bibinfo {author} {P.~Werner}\ and\ \bibinfo {author}
  {T.~Esslinger},\ \emph {\bibinfo {title} {Quantum Simulation Meets
  Nonequilibrium Dynamical Mean-Field Theory: Exploring the Periodically
  Driven, Strongly Correlated Fermi-Hubbard Model}},\ \href
  {\doibase10.1103/physrevlett.123.193602} {\bibfield  {journal} {\bibinfo
  {journal} {Physical Review Letters}\ }\textbf {\bibinfo {volume} {123}},\
  \bibinfo {pages} {193602} (\bibinfo {year} {2019})}\BibitemShut {NoStop}%
\bibitem [{\citenamefont {Liu}\ \emph {et~al.}(2013)\citenamefont {Liu},
  \citenamefont {Hu}, \citenamefont {Wong},\ and\ \citenamefont
  {Xiao}}]{Liu2013}%
  \BibitemOpen
  \bibinfo {author} {Y.-C. Liu}, \bibinfo {author} {Y.-W. Hu}, \bibinfo
  {author} {C.~W. Wong}\ and\ \bibinfo {author} {Y.-F. Xiao},\ \emph {\bibinfo
  {title} {Review of cavity optomechanical cooling}},\ \href
  {\doibase10.1088/1674-1056/22/11/114213} {\bibfield  {journal} {\bibinfo
  {journal} {Chinese Physics B}\ }\textbf {\bibinfo {volume} {22}},\ \bibinfo
  {pages} {114213} (\bibinfo {year} {2013})}\BibitemShut {NoStop}%
\bibitem [{\citenamefont {Marquardt}\ \emph {et~al.}(2008)\citenamefont
  {Marquardt}, \citenamefont {Clerk},\ and\ \citenamefont
  {Girvin}}]{Marquardt2008}%
  \BibitemOpen
  \bibinfo {author} {F.~Marquardt}, \bibinfo {author} {A.~Clerk}\ and\ \bibinfo
  {author} {S.~Girvin},\ \emph {\bibinfo {title} {Quantum theory of
  optomechanical cooling}},\ \href {\doibase10.1080/09500340802454971}
  {\bibfield  {journal} {\bibinfo  {journal} {Journal of Modern Optics}\
  }\textbf {\bibinfo {volume} {55}},\ \bibinfo {pages} {3329} (\bibinfo {year}
  {2008})}\BibitemShut {NoStop}%
\bibitem [{\citenamefont {Weng}\ \emph {et~al.}(2022)\citenamefont {Weng},
  \citenamefont {He}, \citenamefont {Kaszubowska-Anandarajah}, \citenamefont
  {Anandarajah},\ and\ \citenamefont {Kippenberg}}]{Weng2022}%
  \BibitemOpen
  \bibinfo {author} {W.~Weng}, \bibinfo {author} {J.~He}, \bibinfo {author}
  {A.~Kaszubowska-Anandarajah}, \bibinfo {author} {P.~M. Anandarajah}\ and\
  \bibinfo {author} {T.~J. Kippenberg},\ \emph {\bibinfo {title}
  {Microresonator Dissipative Kerr Solitons Synchronized to an Optoelectronic
  Oscillator}},\ \href {\doibase10.1103/physrevapplied.17.024030} {\bibfield
  {journal} {\bibinfo  {journal} {Physical Review Applied}\ }\textbf {\bibinfo
  {volume} {17}},\ \bibinfo {pages} {024030} (\bibinfo {year}
  {2022})}\BibitemShut {NoStop}%
\bibitem [{\citenamefont {Herr}\ \emph {et~al.}(2012)\citenamefont {Herr},
  \citenamefont {Hartinger}, \citenamefont {Riemensberger}, \citenamefont
  {Wang}, \citenamefont {Gavartin}, \citenamefont {Holzwarth}, \citenamefont
  {Gorodetsky},\ and\ \citenamefont {Kippenberg}}]{Herr2012}%
  \BibitemOpen
  \bibinfo {author} {T.~Herr}, \bibinfo {author} {K.~Hartinger}, \bibinfo
  {author} {J.~Riemensberger}, \bibinfo {author} {C.~Y. Wang}, \bibinfo
  {author} {E.~Gavartin}, \bibinfo {author} {R.~Holzwarth}, \bibinfo {author}
  {M.~L. Gorodetsky}\ and\ \bibinfo {author} {T.~J. Kippenberg},\ \emph
  {\bibinfo {title} {Universal formation dynamics and noise of {Kerr}-frequency
  combs in microresonators}},\ \href {\doibase10.1038/nphoton.2012.127}
  {\bibfield  {journal} {\bibinfo  {journal} {Nature Photonics}\ }\textbf
  {\bibinfo {volume} {6}},\ \bibinfo {pages} {480} (\bibinfo {year}
  {2012})}\BibitemShut {NoStop}%
\bibitem [{\citenamefont {Chembo}(2016)}]{Chembo2016}%
  \BibitemOpen
  \bibinfo {author} {Y.~K. Chembo},\ \emph {\bibinfo {title} {Kerr optical
  frequency combs: theory, applications and perspectives}},\ \href
  {\doibase10.1515/nanoph-2016-0013} {\bibfield  {journal} {\bibinfo  {journal}
  {Nanophotonics}\ }\textbf {\bibinfo {volume} {5}},\ \bibinfo {pages} {214}
  (\bibinfo {year} {2016})}\BibitemShut {NoStop}%
\bibitem [{\citenamefont {Lugiato}\ and\ \citenamefont
  {Lefever}(1987)}]{Lugiato1987}%
  \BibitemOpen
  \bibinfo {author} {L.~A. Lugiato}\ and\ \bibinfo {author} {R.~Lefever},\
  \emph {\bibinfo {title} {Spatial Dissipative Structures in Passive Optical
  Systems}},\ \href {\doibase10.1103/physrevlett.58.2209} {\bibfield  {journal}
  {\bibinfo  {journal} {Physical Review Letters}\ }\textbf {\bibinfo {volume}
  {58}},\ \bibinfo {pages} {2209} (\bibinfo {year} {1987})}\BibitemShut
  {NoStop}%
\bibitem [{\citenamefont {del Pino}\ \emph {et~al.}(2023)\citenamefont {del
  Pino}, \citenamefont {Ko{\v{s}}ata},\ and\ \citenamefont
  {Zilberberg}}]{del2023limit}%
  \BibitemOpen
  \bibinfo {author} {J.~del Pino}, \bibinfo {author} {J.~Ko{\v{s}}ata}\ and\
  \bibinfo {author} {O.~Zilberberg},\ \emph {\bibinfo {title} {Limit cycles as
  stationary states of an extended Harmonic Balance ansatz}},\ \href@noop {}
  {\bibfield  {journal} {\bibinfo  {journal} {arXiv preprint arXiv:2308.06092}\
  } (\bibinfo {year} {2023})}\BibitemShut {NoStop}%
\bibitem [{\citenamefont {Flayac}\ and\ \citenamefont
  {Savona}(2017)}]{Flayac2017}%
  \BibitemOpen
  \bibinfo {author} {H.~Flayac}\ and\ \bibinfo {author} {V.~Savona},\ \emph
  {\bibinfo {title} {Unconventional photon blockade}},\ \href
  {\doibase10.1103/physreva.96.053810} {\bibfield  {journal} {\bibinfo
  {journal} {Physical Review A}\ }\textbf {\bibinfo {volume} {96}},\ \bibinfo
  {pages} {053810} (\bibinfo {year} {2017})}\BibitemShut {NoStop}%
\bibitem [{\citenamefont {Ling}\ \emph {et~al.}(2023)\citenamefont {Ling},
  \citenamefont {Qvarfort},\ and\ \citenamefont {Mintert}}]{Ling2023}%
  \BibitemOpen
  \bibinfo {author} {Y.~Ling}, \bibinfo {author} {S.~Qvarfort}\ and\ \bibinfo
  {author} {F.~Mintert},\ \emph {\bibinfo {title} {Fast optomechanical photon
  blockade}},\ \href {\doibase10.1103/physrevresearch.5.023148} {\bibfield
  {journal} {\bibinfo  {journal} {Physical Review Research}\ }\textbf {\bibinfo
  {volume} {5}},\ \bibinfo {pages} {023148} (\bibinfo {year}
  {2023})}\BibitemShut {NoStop}%
\bibitem [{\citenamefont {Chen}\ \emph {et~al.}(2022)\citenamefont {Chen},
  \citenamefont {Tang}, \citenamefont {Tang}, \citenamefont {Wu},\ and\
  \citenamefont {Xia}}]{Chen2022}%
  \BibitemOpen
  \bibinfo {author} {M.~Chen}, \bibinfo {author} {J.~Tang}, \bibinfo {author}
  {L.~Tang}, \bibinfo {author} {H.~Wu}\ and\ \bibinfo {author} {K.~Xia},\ \emph
  {\bibinfo {title} {Photon blockade and single-photon generation with multiple
  quantum emitters}},\ \href {\doibase10.1103/physrevresearch.4.033083}
  {\bibfield  {journal} {\bibinfo  {journal} {Physical Review Research}\
  }\textbf {\bibinfo {volume} {4}},\ \bibinfo {pages} {033083} (\bibinfo {year}
  {2022})}\BibitemShut {NoStop}%
\bibitem [{\citenamefont {Xiao}\ \emph {et~al.}(2023)\citenamefont {Xiao},
  \citenamefont {Venkatraman}, \citenamefont {Cortiñas}, \citenamefont
  {Chowdhury},\ and\ \citenamefont {Devoret}}]{Xiao2023}%
  \BibitemOpen
  \bibinfo {author} {X.~Xiao}, \bibinfo {author} {J.~Venkatraman}, \bibinfo
  {author} {R.~G. Cortiñas}, \bibinfo {author} {S.~Chowdhury}\ and\ \bibinfo
  {author} {M.~H. Devoret},\ \href {\doibase10.48550/ARXIV.2304.13656} {\emph
  {\bibinfo {title} {A diagrammatic method to compute the effective Hamiltonian
  of driven nonlinear oscillators}}} (\bibinfo {year} {2023})\BibitemShut
  {NoStop}%
\bibitem [{\citenamefont {Biondi}\ \emph {et~al.}(2017)\citenamefont {Biondi},
  \citenamefont {Blatter}, \citenamefont {Türeci},\ and\ \citenamefont
  {Schmidt}}]{Biondi2017}%
  \BibitemOpen
  \bibinfo {author} {M.~Biondi}, \bibinfo {author} {G.~Blatter}, \bibinfo
  {author} {H.~E. Türeci}\ and\ \bibinfo {author} {S.~Schmidt},\ \emph
  {\bibinfo {title} {Nonequilibrium gas-liquid transition in the
  driven-dissipative photonic lattice}},\ \href
  {\doibase10.1103/PhysRevA.96.043809} {\bibfield  {journal} {\bibinfo
  {journal} {Physical Review A}\ }\textbf {\bibinfo {volume} {96}},\ \bibinfo
  {pages} {043809} (\bibinfo {year} {2017})}\BibitemShut {NoStop}%
\bibitem [{\citenamefont {Casteels}\ \emph {et~al.}(2017)\citenamefont
  {Casteels}, \citenamefont {Fazio},\ and\ \citenamefont
  {Ciuti}}]{Casteels2017}%
  \BibitemOpen
  \bibinfo {author} {W.~Casteels}, \bibinfo {author} {R.~Fazio}\ and\ \bibinfo
  {author} {C.~Ciuti},\ \emph {\bibinfo {title} {Critical dynamical properties
  of a first-order dissipative phase transition}},\ \href
  {\doibase10.1103/PhysRevA.95.012128} {\bibfield  {journal} {\bibinfo
  {journal} {Physical Review A}\ }\textbf {\bibinfo {volume} {95}},\ \bibinfo
  {pages} {012128} (\bibinfo {year} {2017})}\BibitemShut {NoStop}%
\bibitem [{\citenamefont {Košata}\ \emph {et~al.}(2022)\citenamefont
  {Košata}, \citenamefont {Leuch}, \citenamefont {Kästli},\ and\
  \citenamefont {Zilberberg}}]{Kosata2022}%
  \BibitemOpen
  \bibinfo {author} {J.~Košata}, \bibinfo {author} {A.~Leuch}, \bibinfo
  {author} {T.~Kästli}\ and\ \bibinfo {author} {O.~Zilberberg},\ \emph
  {\bibinfo {title} {Fixing the rotating-wave approximation for strongly
  detuned quantum oscillators}},\ \href
  {\doibase10.1103/PhysRevResearch.4.033177} {\bibfield  {journal} {\bibinfo
  {journal} {Physical Review Research}\ }\textbf {\bibinfo {volume} {4}},\
  \bibinfo {pages} {033177} (\bibinfo {year} {2022})}\BibitemShut {NoStop}%
\bibitem [{\citenamefont {Lifshitz}\ and\ \citenamefont
  {Cross}(2008)}]{Lifshitz2008}%
  \BibitemOpen
  \bibinfo {author} {R.~Lifshitz}\ and\ \bibinfo {author} {M.~C. Cross},\ \emph
  {\bibinfo {title} {Nonlinear Dynamics of Nanomechanical and Micromechanical
  Resonators}},\ in\ \href {\doibase10.1002/9783527626359.ch1} {\emph {\bibinfo
  {booktitle} {Reviews of Nonlinear Dynamics and Complexity}}}\ (\bibinfo
  {publisher} {John Wiley \& Sons, Ltd},\ \bibinfo {year} {2008})\
  Chap.~\bibinfo {chapter} {1}, pp.\ \bibinfo {pages} {1--52}\BibitemShut
  {NoStop}%
\bibitem [{sup()}]{supmat}%
  \BibitemOpen
  \href@noop {} {}\bibinfo {note} {See Supplemental Material.}\BibitemShut
  {Stop}%
\bibitem [{Note5()}]{Note5}%
  \BibitemOpen
  \bibinfo {note} {Here, we follow naming convention similar the one use in
  Ref.~\cite {Eckardt2015}, i.e., the effective Hamiltonian is the
  time-independent Hamiltonian which represent the stroboscopic dynamics of the
  system independent of the Floquet gauge --- in contrast to the Floquet
  Hamiltonian.}\BibitemShut {Stop}%
\bibitem [{Note10()}]{Note10}%
  \BibitemOpen
  \bibinfo {note} {Applying the quantum-to-classical limit to the unitary
  rotating frame transformation $\protect \mathcal {U}_b(t)$ yields the same
  rotation transformation used in the classical framework. This is not the case
  when working in the $a$-basis and the rotating frame $\protect \mathcal
  {U}_a(t)$~\cite {supmat}. Specifically, we can write $\protect \hat {a} =
  \protect \hat {S}^\dagger \protect \hat {b}\protect \hat {S}$, with $\protect
  \hat {S}=\exp (z/2(\protect \co ^2+\protect \ao ^2))$, where the parameter
  $z$ controls the symplectic transformation and is a function of $\omega _0$
  and $\omega $. Here, $\protect \ao = \mu \protect \ao [b]-\nu \protect \co
  [b]$, with $\mu =1/2(\protect \sqrt {\omega /\omega _0}+\protect \sqrt
  {\omega _0/\omega })$, and $\nu =1/2(\protect \sqrt {\omega /\omega
  _0}-\protect \sqrt {\omega _0/\omega })$. The parameter $z$ is defined by
  $\mu =\cosh (|z|)$ and $\nu =z/|z|\sinh (|z|)$.}\BibitemShut {Stop}%
\bibitem [{\citenamefont {Anikin}\ \emph {et~al.}(2021)\citenamefont {Anikin},
  \citenamefont {Maslova}, \citenamefont {Gippius},\ and\ \citenamefont
  {Sokolov}}]{Anikin2021}%
  \BibitemOpen
  \bibinfo {author} {E.~V. Anikin}, \bibinfo {author} {N.~S. Maslova}, \bibinfo
  {author} {N.~A. Gippius}\ and\ \bibinfo {author} {I.~M. Sokolov},\ \emph
  {\bibinfo {title} {Multiphoton resonance in a driven Kerr oscillator in the
  presence of high-order nonlinearities}},\ \href
  {\doibase10.1103/physreva.104.053106} {\bibfield  {journal} {\bibinfo
  {journal} {Physical Review A}\ }\textbf {\bibinfo {volume} {104}},\ \bibinfo
  {pages} {053106} (\bibinfo {year} {2021})}\BibitemShut {NoStop}%
\bibitem [{\citenamefont {Ruiz}\ \emph {et~al.}(2023)\citenamefont {Ruiz},
  \citenamefont {Gautier}, \citenamefont {Guillaud},\ and\ \citenamefont
  {Mirrahimi}}]{Ruiz2023}%
  \BibitemOpen
  \bibinfo {author} {D.~Ruiz}, \bibinfo {author} {R.~Gautier}, \bibinfo
  {author} {J.~Guillaud}\ and\ \bibinfo {author} {M.~Mirrahimi},\ \emph
  {\bibinfo {title} {Two-photon driven Kerr quantum oscillator with multiple
  spectral degeneracies}},\ \href {\doibase10.1103/physreva.107.042407}
  {\bibfield  {journal} {\bibinfo  {journal} {Physical Review A}\ }\textbf
  {\bibinfo {volume} {107}},\ \bibinfo {pages} {042407} (\bibinfo {year}
  {2023})}\BibitemShut {NoStop}%
\bibitem [{\citenamefont {Winkel}\ \emph {et~al.}(2020)\citenamefont {Winkel},
  \citenamefont {Borisov}, \citenamefont {Grünhaupt}, \citenamefont {Rieger},
  \citenamefont {Spiecker}, \citenamefont {Valenti}, \citenamefont {Ustinov},
  \citenamefont {Wernsdorfer},\ and\ \citenamefont {Pop}}]{Winkel2020}%
  \BibitemOpen
  \bibinfo {author} {P.~Winkel}, \bibinfo {author} {K.~Borisov}, \bibinfo
  {author} {L.~Grünhaupt}, \bibinfo {author} {D.~Rieger}, \bibinfo {author}
  {M.~Spiecker}, \bibinfo {author} {F.~Valenti}, \bibinfo {author} {A.~V.
  Ustinov}, \bibinfo {author} {W.~Wernsdorfer}\ and\ \bibinfo {author} {I.~M.
  Pop},\ \emph {\bibinfo {title} {Implementation of a {Transmon} {Qubit}
  {Using} {Superconducting} {Granular} {Aluminum}}},\ \href
  {\doibase10.1103/PhysRevX.10.031032} {\bibfield  {journal} {\bibinfo
  {journal} {Physical Review X}\ }\textbf {\bibinfo {volume} {10}},\ \bibinfo
  {pages} {031032} (\bibinfo {year} {2020})}\BibitemShut {NoStop}%
\bibitem [{\citenamefont {Drummond}\ and\ \citenamefont
  {Walls}(1980)}]{Drummond1980}%
  \BibitemOpen
  \bibinfo {author} {P.~D. Drummond}\ and\ \bibinfo {author} {D.~F. Walls},\
  \emph {\bibinfo {title} {Quantum theory of optical bistability. I. Nonlinear
  polarisability model}},\ \href {\doibase10.1088/0305-4470/13/2/034}
  {\bibfield  {journal} {\bibinfo  {journal} {Journal of Physics A:
  Mathematical and General}\ }\textbf {\bibinfo {volume} {13}},\ \bibinfo
  {pages} {725} (\bibinfo {year} {1980})}\BibitemShut {NoStop}%
\bibitem [{\citenamefont {Bartolo}\ \emph {et~al.}(2016)\citenamefont
  {Bartolo}, \citenamefont {Minganti}, \citenamefont {Casteels},\ and\
  \citenamefont {Ciuti}}]{Bartolo2016}%
  \BibitemOpen
  \bibinfo {author} {N.~Bartolo}, \bibinfo {author} {F.~Minganti}, \bibinfo
  {author} {W.~Casteels}\ and\ \bibinfo {author} {C.~Ciuti},\ \emph {\bibinfo
  {title} {Exact steady state of a {Kerr} resonator with one- and two-photon
  driving and dissipation: {Controllable} {Wigner}-function multimodality and
  dissipative phase transitions}},\ \href {\doibase10.1103/PhysRevA.94.033841}
  {\bibfield  {journal} {\bibinfo  {journal} {Physical Review A}\ }\textbf
  {\bibinfo {volume} {94}},\ \bibinfo {pages} {033841} (\bibinfo {year}
  {2016})}\BibitemShut {NoStop}%
\bibitem [{\citenamefont {Roberts}\ and\ \citenamefont
  {Clerk}(2020)}]{Roberts2020}%
  \BibitemOpen
  \bibinfo {author} {D.~Roberts}\ and\ \bibinfo {author} {A.~A. Clerk},\ \emph
  {\bibinfo {title} {Driven-Dissipative Quantum Kerr Resonators: New Exact
  Solutions, Photon Blockade and Quantum Bistability}},\ \href
  {\doibase10.1103/physrevx.10.021022} {\bibfield  {journal} {\bibinfo
  {journal} {Physical Review X}\ }\textbf {\bibinfo {volume} {10}},\ \bibinfo
  {pages} {021022} (\bibinfo {year} {2020})}\BibitemShut {NoStop}%
\bibitem [{\citenamefont {Meade}(1982)}]{Meade1982}%
  \BibitemOpen
  \bibinfo {author} {M.~L. Meade},\ \emph {\bibinfo {title} {Advances in
  lock-in amplifiers}},\ \href {\doibase10.1088/0022-3735/15/4/001} {\bibfield
  {journal} {\bibinfo  {journal} {Journal of Physics E: Scientific
  Instruments}\ }\textbf {\bibinfo {volume} {15}},\ \bibinfo {pages} {395}
  (\bibinfo {year} {1982})}\BibitemShut {NoStop}%
\bibitem [{\citenamefont {Breuer}(2007)}]{Breuer2007}%
  \BibitemOpen
  \bibinfo {author} {H.-P. Breuer},\ \href
  {\doibase10.1093/acprof:oso/9780199213900.001.0001} {\emph {\bibinfo {title}
  {The Theory of Open Quantum Systems}}}\ (\bibinfo  {publisher} {Clarendon},\
  \bibinfo {address} {Oxford},\ \bibinfo {year} {2007})\BibitemShut {NoStop}%
\bibitem [{\citenamefont {Gardiner}\ and\ \citenamefont
  {Zoller}(2000)}]{Gardiner2000}%
  \BibitemOpen
  \bibinfo {author} {C.~Gardiner}\ and\ \bibinfo {author} {P.~Zoller},\ \href
  {https://books.google.ch/books?id=4bJ6MgEACAAJ} {\emph {\bibinfo {title}
  {Quantum Noise: A Handbook of Markovian and Non-Markovian Quantum Stochastic
  Methods with Applications to Quantum Optics}}},\ Springer series in
  synergetics\ (\bibinfo  {publisher} {Springer Berlin Heidelberg},\ \bibinfo
  {year} {2000})\BibitemShut {NoStop}%
\end{thebibliography}%

\end{document}